\documentclass{rspublicmod}

\usepackage{graphicx}
\usepackage{mathrsfs}
\usepackage{lineno}
\usepackage{hyperref}

\usepackage{caption}
\usepackage{upgreek}

\newcommand{\figlabel}[1]{{\itshape (#1)}}
\newcommand{\leavethisout}[1]{}
\newcommand{\etal}{{\itshape et~al.}}
\newcommand{\myee}[2]{\mbox{${#1}\times 10^{#2}$}}
\newcommand{\mymc}[1]{\multicolumn{2}{c}{#1}}
\newcommand{\degC}{${}^\circ$C}
\newcommand{\degK}{${}^\circ$K}
\newcommand{\cellarea}{A}
\newcommand{\hydcond}{\mathscr{L}}

\usepackage[fixpdftex]{xcolor}
\newcommand{\myrevision}[1]{{\color{red}#1}}
\renewcommand{\myrevision}[1]{#1}

\begin{document}


\title[Multiscale model for sap exudation]{Multiscale model of a
  freeze--thaw process for tree sap exudation}

\author[I. Graf, M. Ceseri and J. M. Stockie]{%
  Isabell Graf${}^{1}$, 
  Maurizio Ceseri${}^{2}$,  
  and John M. Stockie${}^{1,*}$}

\affiliation{%
  $^{1}$Department of Mathematics, Simon Fraser University,   
  8888 University Drive, \\ 
  Burnaby, British Columbia, V5A 1S6, Canada\\
  $^{2}$Istituto per le Applicazioni del Calcolo `Mauro Picone', 
  via dei Taurini 19, \\
  Consiglio Nazionale delle Ricerche, Rome, 00185, Italy\\
  $^\ast$Author for correspondence: John M. Stockie
  (stockie@math.sfu.ca)
} 

\label{firstpage}

\maketitle

\begin{abstract}{%
    Tree sap exudation; sugar maple; multiphase flow and transport;
    phase change; differential equations; periodic homogenization 
  }
  Sap transport in trees has long fascinated scientists, and a vast
  literature exists on experimental and modelling studies of trees
  during the growing season when large negative stem pressures are
  generated by transpiration from leaves.  Much less attention has been
  paid to winter months when trees are largely dormant but nonetheless
  continue to exhibit interesting flow behaviour.  A prime example is
  sap exudation, which refers to the peculiar ability of sugar maple
  (\emph{Acer saccharum}) and related species to generate positive stem
  pressure while in a leafless state.  Experiments demonstrate that
  ambient temperatures must oscillate about the freezing point before
  significantly heightened stem pressures are observed, but the precise
  causes of exudation remain unresolved.  The prevailing hypothesis
  attributes exudation to a physical process combining freeze--thaw and
  osmosis, which has some support from experimental studies but remains
  a subject of active debate.  We address this knowledge gap by
  developing the first mathematical model for exudation, while also
  introducing several essential modifications to this hypothesis.  We
  derive a multiscale model consisting of a nonlinear system of
  differential equations governing phase change and transport within
  wood cells, coupled to a suitably homogenized equation for temperature
  on the macroscale.  Numerical simulations yield stem pressures that
  are consistent with experiments and provide convincing evidence that a
  purely physical mechanism is capable of capturing exudation.
\end{abstract}

\section{Introduction}
\label{sec:intro}

The study of tree sap flow has a long history that has given rise over
time to the concept of the hydraulic architecture of
trees~\cite{tyree-zimmermann-2002}. 
Despite the extensive literature on this subject, several aspects of sap
transport remain controversial, including the cohesion-tension theory of
sap ascent \cite{angeles-etal-2004, tyree-2003, zimmermann-etal-2004};
embolism formation and recovery
\cite{meinzer-clearwater-goldstein-2001, zwienicki-holbrook-2009}, 
which is ubiquitous in species subject to drought- or freezing-induced
stresses; and sap exudation in maple and related species such as walnut,
butternut and birch~\cite{cirelli-etal-2008}.  Furthermore, there is a
great deal of current interest in the possible effects of recent
changes in weather patterns on both individual trees and forest
ecosystems~\cite{beckage-etal-2008, groffman-etal-2012}, and their
connections with sap hydraulics 
\cite{sperry-love-2015}.  The problems just described involve complex
interactions between sap flow and other phenomena such as nutrient
transport, photosynthesis, soil physics, atmospheric dynamics, cell
growth, etc.  Despite the extensive work to date on mathematical and
computational modelling of trees and their interactions with the
environment, 
many open questions remain that can only be addressed by
considering sap flow coupled with other processes and building
integrated models that connect flow and structure at different spatial
scales and levels of organization~\cite{kim-park-hwang-2014}.
%



Sugar maple is a keystone species in the forests of central and eastern
North America~\cite{minorsky-2003b} and so is worthy of special
attention.  Members of the maple family are distinguished from 
other hardwoods by a number of unusual structural and functional
features that allow them to exude sap during
winter~\cite{milburn-omalley-1984, tyree-1983}, to generate unusually
high rates of nitrification~\cite{lovett-mitchell-2004}, or to recover
from freeze-induced
embolism~\cite{sperry-etal-1988, yang-tyree-1992}.  The potential impacts
of climate change on maple have also attracted recent
attention~\cite{lovett-mitchell-2004, minorsky-2003b},\
motivated by the economic importance of the maple syrup industry, not to
mention maple's high timber value.  In particular, maple sap yields are
sensitive to even small variations in temperature or snow cover during
the harvest season, so that recent unusual weather patterns underscore
the importance of developing a better understanding of 
the effects of local environmental conditions on sap
flow~\cite{karl-melillo-peterson-2009, reynolds-2010}.

Hundreds of scientific papers have addressed the phenomenon of sap
exudation during winter when maple trees \myrevision{are leafless and
  yet still exhibit pressure variations that range over
  150--180\3kPa~\cite{ameglio-etal-2001, cirelli-etal-2008,
    cortes-sinclair-1985, tyree-1983}.} 
However, the precise mechanism driving the generation of heightened
exudation pressure is still not fully
understood~\cite{tyree-zimmermann-2002}.  The first systematic study
appeared in an 1860 article by Sachs~\cite{sachs-1860}, who attributed
exudation pressure to thermal expansion of gas within sapwood or
\emph{xylem}.  The next major advance in understanding followed from the
exhaustive study of Wiegand~\cite{wiegand-1906}, who found to the
contrary that thermal expansion of gas, water or wood has minimal impact
on exudation.  Instead, Wiegand proposed a vitalistic or `living cell'
hypothesis wherein sugar is released into the sap by some cellular
activity, which gives rise to elevated pressure from osmotic gradients
across selectively-permeable membranes separating wood cells.
Subsequently, this osmotic mechanism figured prominently in the
literature, although experimental studies have continued to yield
conflicting results that in turn stimulated development of new theories
advocating alternate (bio-)physical mechanisms.  For example, some
authors continued to support the thermal expansion
hypothesis~\cite{merwin-lyon-1909}, while others advocated the various
roles of gas dissolution~\cite{johnson-1945}, cryostatic suction due to
freezing~\cite{stevens-eggert-1945}, or temperature-induced changes in
bark thickness~\cite{marvin-1949}.  More recent studies have led to a
new understanding of exudation as a physical process deriving from a
combination of freezing and thawing of sap~\cite{milburn-omalley-1984}
with osmosis~\cite{tyree-1995}.  Although some experimental evidence
supports this hypothesis~\cite{cirelli-etal-2008} the precise mechanisms
behind the exudation phenomenon is still not fully understood.

We aim to resolve this long-standing open question by developing the
first mathematical model for the freeze--thaw process in maple.  We
uncover the essential role played by two physical mechanisms 
whose significance has not yet been recognized -- namely,
root water uptake and freezing point depression due to sap sugar
content.  Using numerical simulations of repeated freeze--thaw
cycles, we obtain computed exudation pressures that are consistent with
experimental results.

Although the focus of this paper is on developing a complete and
physically consistent model for sap exudation, our results also have
more far-reaching consequences.  This work affords new insights into the
complex multi-physics processes occurring in trees and also provides a
framework for studying other practical questions of importance to tree
physiologists and maple syrup producers.  Our model also provides a
platform for studying related phenomena such as embolism that occur in a
much broader range of tree species, as well as evaluating the response
of trees to changes in environmental variables such as temperature and
soil moisture arising from various climate change scenarios.

\section{Physical mechanism for sap exudation}
\label{sec:physical}

\subsection{Milburn and O'Malley's hypothesis} 
\begin{figure}
  \centering
  \begin{tabular}{lll}
    \figlabel{a}  & \qquad\qquad &
    \myrevision{\figlabel{b}} \\
    %
    %
    {\setlength{\unitlength}{0.33\textwidth}
      \begin{picture}(1,1)
        \put(0,0){\includegraphics[width=0.33\textwidth]{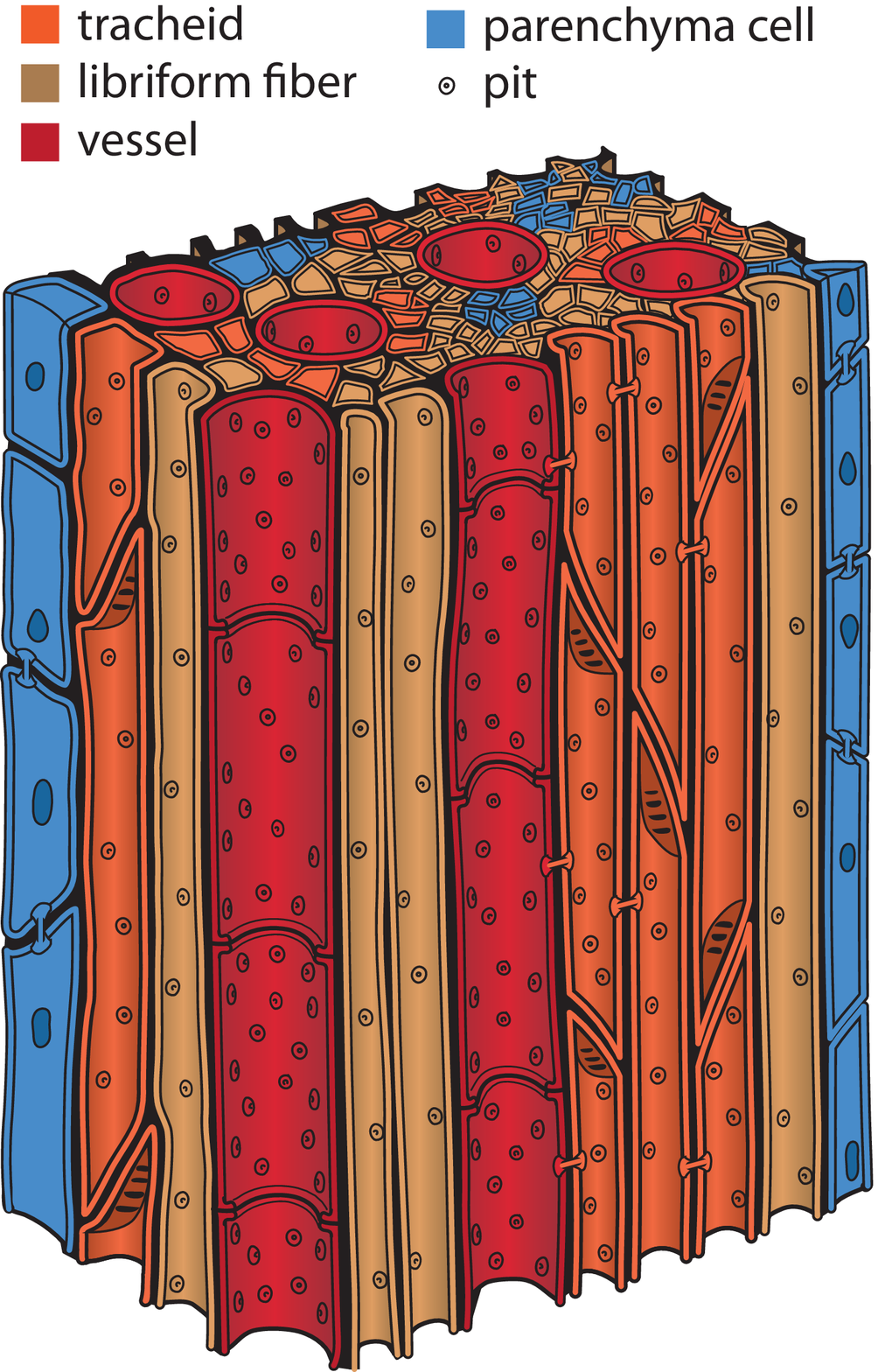}}
        \thicklines
        \linethickness{0.5mm}
        \put(0.7,0.02){\line(1,0){0.3}}
        \put(0.7,0.0){\line(0,1){0.04}}
        \put(1.0,0.0){\line(0,1){0.04}}
        \put(0.75,0.05){\footnotesize\sffamily 100\,$\upmu$m} 
      \end{picture}
    }
    & &
    \includegraphics[width=0.30\textwidth]{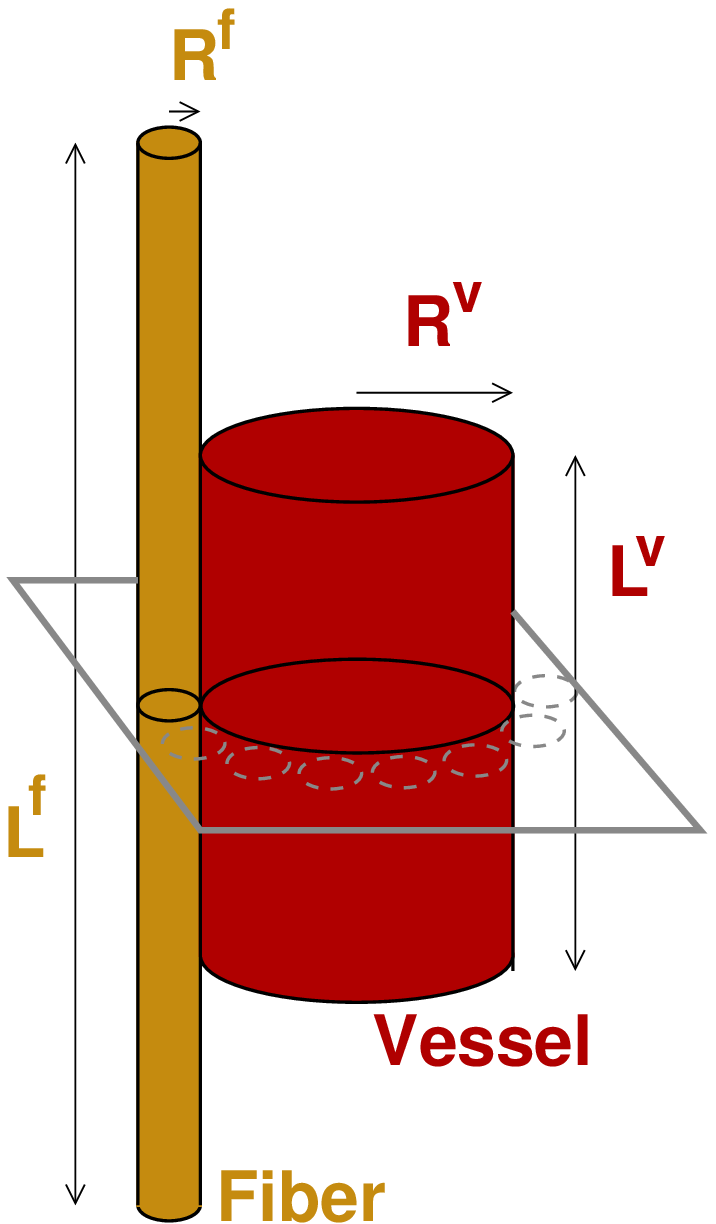}
  \end{tabular}
  \caption{{\bf Xylem microstructure.}  \figlabel{a} Cross-sectional
    view of hardwood xylem, showing tracheids connected hydraulically to
    vessels and other tracheids via paired pits.  Fibers appear similar
    to tracheids except that they have fewer pits, most of which are
    blind or unpaired.  \myrevision{The parenchyma are living cells
      whose main role is carbohydrate storage and so they
      are ignored here.}
    \myrevision{\figlabel{b} A fiber--vessel pair approximated as
      circular cylinders, showing typical dimensions of the fiber
      (length $L^f=1.0\times 10^{-3}\3$m and radius $R^f=3.5\times
      10^{-6}\3$) and vessel ($L^v=5.0\times 10^{-4}\3$m and
      $R^v=2.0\times 10^{-5}\3$m).  The model domain corresponds to
      the horizontal cross-section through the middle of the diagram.}}
  \label{fig:sapwood}
\end{figure}

%
Experimental work up to the 1980's demonstrated that no single physical
mechanism is capable of capturing measured winter stem
pressures~\cite{marvin-1968}, and sap exudation remained an unsolved
puzzle until the ground-breaking study of Milburn and
O'Malley~\cite{milburn-omalley-1984}.
They proposed a physical mechanism based on freezing and
thawing of sap, motivated by the unique structural characteristic of
xylem in maple (and related trees) that a significant proportion
of the \emph{libriform fibers} (or simply fibers) \myrevision{are
  primarily gas-filled rather than being liquid-filled} as in most other
hardwood species~\cite{wiegand-1906}.  This peculiar feature of the fibers should
be contrasted with the two other cell types that play an active role in
sap transport -- \emph{vessels} and \emph{tracheids} -- which are mostly
sap-filled and are connected hydraulically to each other via {paired
  pits} (see figure~\ref{fig:sapwood}a).  Indeed, recent
experiments~\cite{cirelli-etal-2008, sano-etal-2011}
\myrevision{suggest} that fibers are essentially non-conductive
\myrevision{in comparison with other xylem elements because they 
  lack end-to-end cell connections and their lateral} walls
contain mostly unpaired (blind) pits that are smaller and fewer in
number than the more conductive vessels and tracheids.


\begin{figure}
  \centering
  \includegraphics[width=0.70\textwidth]{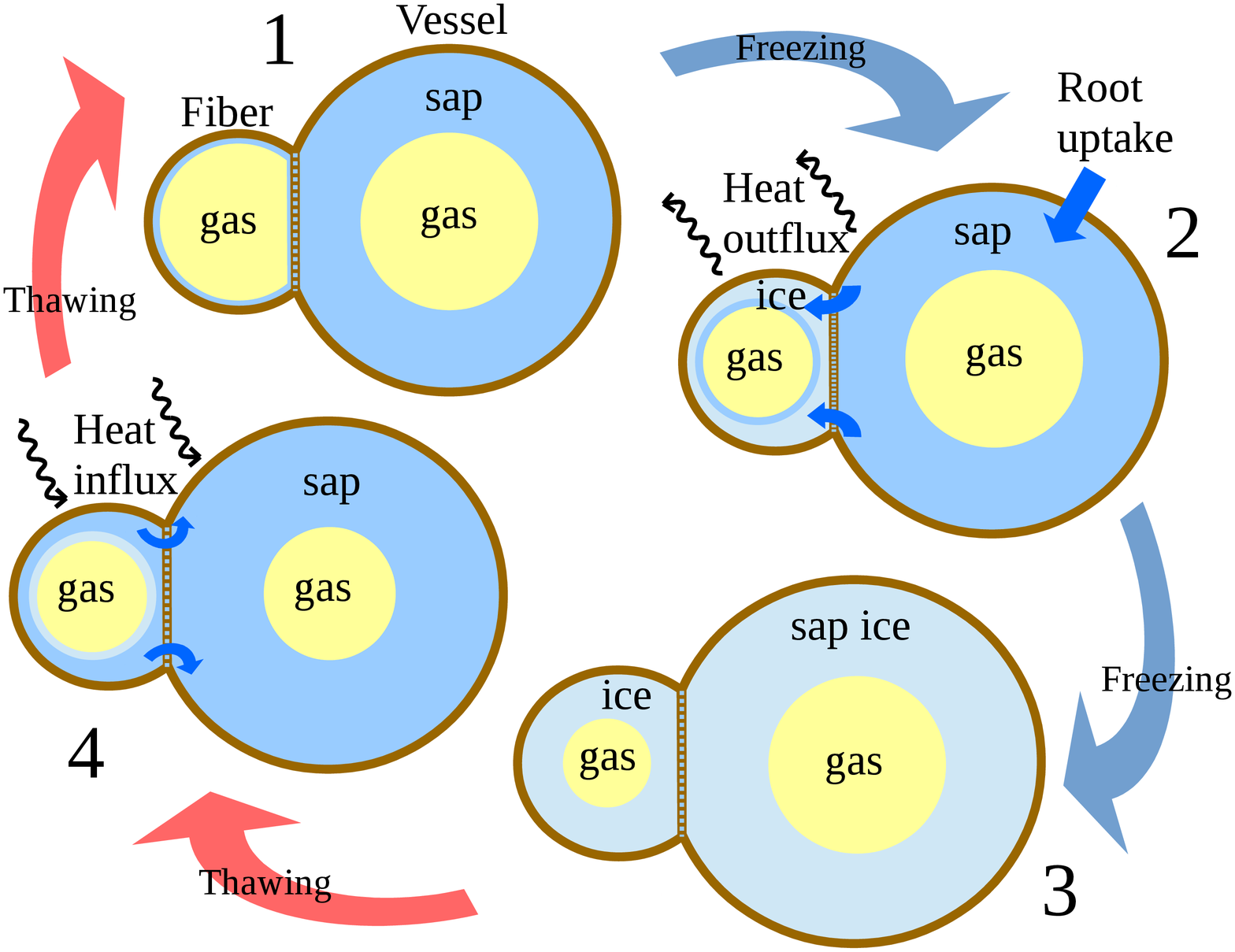}
  \caption{{\bf Stages in the freeze--thaw cycle.}  \myrevision{Various
      stages in the freeze--thaw cycle are depicted within an adjacent
      fiber--vessel pair.}  Stages 
    1$\rightarrow$2$\rightarrow$3 depict the freezing process:
    when temperature drops, an ice layer grows on the inner
    wall of the gas-filled fiber as water is drawn via cryostatic
    suction through the porous cell wall.  Stages
    3$\rightarrow$4$\rightarrow$1 depict the reverse process as
    temperature rises.  Note the reversed order of phase interfaces
    inside the fiber between stages 2 and 4.  \myrevision{The blue arrows
      denote water transport either through the fiber--vessel wall
      or between roots and vessels.
    }}
  \label{fig:cycle-stages}
\end{figure}

\myrevision{Milburn and O'Malley focused on the dynamics of a single
  fiber--vessel pair as pictured in figure~\ref{fig:sapwood}b, and
  ignored other xylem elements such as parenchyma and ray cells.
  Whereas fibers had previously been thought} to play a purely
structural role, Milburn and O'Malley proposed that as temperature falls
below freezing, liquid from the vessel is drawn by cryostatic suction
through the porous fiber--vessel wall to freeze on the inside of the
fiber (figure~\ref{fig:cycle-stages}, stages 1-2-3).  As a result, any
gas contained within the fiber is compressed and acts as a pressure
reservoir.  When temperature subsequently rises and ice thaws, the
process reverses and the compressed gas bubble forces melt-water into
the vessel, thereby re-pressurizing the vessel sap
(figure~\ref{fig:cycle-stages}, stages 3-4-1).

\subsection{Tyree's modified hypothesis, with gas dissolution and osmosis}
%
This freeze--thaw hypothesis was critically evaluated by
Tyree~\cite{tyree-1995}, who \myrevision{proposed a modified hypothesis
  featuring two important additions.}  First, he recognized that gas
under pressure will dissolve within an adjacent
liquid~\cite{keller-1964} and that pressures encountered in maple xylem
are high enough that any bubbles should dissolve completely given
sufficient time~\cite{tyree-yang-1992}. Therefore, some additional
mechanism is required to sustain gas bubbles in the fibers.  Tyree's
second observation was that measured xylem pressures depend strongly on
sugar concentration in the vessel sap, 80\%\ of which derives from
sucrose~\cite{cortes-sinclair-1985},
which led him to conclude that sucrose is required for exudation.
\myrevision{He recognized that although the axial conductivity of fibers
  is negligible in comparison with vessels and tracheids, the lignified
  cellulose making up secondary cell walls should admit a small radial
  conductivity.  He then hypothesized that the fiber--vessel wall forms
  an osmotic barrier that allows water to penetrate but not the larger
  sucrose molecules}.  Consequently, an \myrevision{additional} osmotic
pressure difference exists between the sweet vessel sap and pure fiber
water, which he argued is responsible for preventing fiber gas
bubbles from completely dissolving.

Tyree's \myrevision{modified freeze--thaw hypothesis} includes water
phase transitions, gas dissolution and osmosis, and is currently the
prevailing hypothesis for sap exudation~\cite{tyree-1995}.  It depends
strongly on the existence of a hydraulically isolated system of fibers
and a selectively permeable fiber--vessel wall, both of which have since
been confirmed experimentally~\cite{cirelli-etal-2008}.
Although this evidence is compelling, there has been no attempt yet to
model this process mathematically (except for a related process without
phase change in the context of embolism recovery in
maple~\cite{yang-tyree-1992}) and so it remains unclear whether this
physical description is capable of capturing exudation.

\subsection{Three essential physical mechanisms} 
%
Before proceeding further, we extend the freeze--thaw hypothesis just
described by incorporating three additional mechanisms:
\begin{description}
\item[\normalfont\emph{Gas bubbles in the vessel:}] Sap (like water) is
  an incompressible fluid so that in the rigid, closed vessel network of
  a leafless tree there is no mechanism for fiber--vessel
  \myrevision{mass} transfer if vessels are completely saturated with
  sap.  However, the existence of gas bubbles within vessels is
  well-documented in maple~\cite{perkins-vandenberg-2009, wiegand-1906}
  and other hardwood species~\cite{tyree-1995}.  Even if xylem pressures
  were high enough to dissolve such bubbles, gas would eventually be
  forced out of solution upon freezing and so at least a transient
  presence of gas bubbles is unavoidable.
  Therefore, introducing a gas phase in the vessels provides a plausible
  mechanism for fiber--vessel pressure exchange.
  
\item[\normalfont\emph{Sap freezing point depression (FPD):}] Sap
  contains dissolved sugars and hence experiences a reduced freezing
  point compared to pure water according to Blagden's
  Law~\cite{cavenderbares-2005}, $\Delta T_{fpd} = K_b \,C_s/\rho_w$,
  where $K_b$ is the cryoscopic constant, $C_s$ is sugar concentration,
  and $\rho_w$ is water density.  For example, sap containing 3\%\
  sucrose \myrevision{by mass} experiences a FPD of $\Delta T_{fpd}
  \approx 0.162\3$\degK.  Although this temperature difference may
  appear insignificant, we will see that it is actually large when
  considered on the scale of individual cells, and indeed is sufficient
  to account for the existence of ice in fibers while sap in adjacent
  vessels remains in liquid form. \myrevision{This partitioning of ice
    and liquid in neighbouring fiber--vessel pairs induces cryostatic
    suction that draws liquid out of the vessel to form ice on the inner
    fiber wall.}

\item[\normalfont\emph{Root water uptake during freezing:}]
  \myrevision{No previous hypothesis for sap exudation explicitly
    considers the role of root water uptake.  Furthermore, several}
  studies suggest that root pressure in maple has a negligible effect on
  exudation~\cite{ameglio-etal-2001, kramer-boyer-1995}.
  Nonetheless, it is well-known that during winter months trees can draw
  water from the roots if soil temperatures are high
  enough~\cite{stevens-eggert-1945, tyree-zimmermann-2002}, which can be
  caused by an insulating snow
  cover~\cite{robitaille-boutin-lachance-1995}.  Recent experiments on
  maple saplings~\cite{brown-2013, perkins-vandenberg-patent-2015} have
  provided the first direct evidence that root water uptake occurs in
  maple during winter while exudation is underway.
\end{description}
Our aim is now to incorporate these three modifications into a model of
the freeze--thaw process outlined previously, and then demonstrate that
the resulting equations are capable of reproducing observed behaviours.

\section{Mathematical formulation} 
\label{sec:model}

\subsection{Outline of the modelling approach}

The freeze--thaw mechanism outlined in the previous section 
involves processes operating on two distinct spatial scales: the
\emph{microscale} corresponding to individual wood cells with 
dimensions ranging from 10--100 microns; and the \emph{macroscale}
corresponding to the tree stem with diameter tens of centimetres. The
derivation of our mathematical model for sap exudation therefore divides 
naturally over these two scales.  Firstly, we develop microscale
equations that capture cell-level processes within libriform fibers and
vessels, combining the dynamics of freezing, thawing, gas
dissolution, osmotic pressure, heat transport, and porous flow through
the fiber--vessel wall.  Secondly, we consider heat transport in the
entire tree stem and apply periodic homogenization to derive an equation
for the macroscale temperature that incorporates microscale cellular
processes via appropriately defined transport coefficients and source
terms.  We proceed as follows:
\begin{itemize}
\item Start from an existing microscale model for the thawing half of
  the freeze--thaw process~\cite{ceseri-stockie-2013, graf-stockie-2014}
  in which a 2D periodic microstructure is assembled from copies of a
  \emph{reference cell} $Y$ containing a single fiber and vessel (see
  figure~\ref{fig:homo}a).  The fiber is placed at the centre of $Y$
  (where the dashed line denotes the fiber--vessel wall) and the
  remainder of the reference cell corresponds to the vessel.  This
  choice of geometry is a mathematical idealization that captures the
  volumes of the fiber and vessel compartments, but is not intended to
  accurately represent the actual layout of wood cells.  The governing
  equations consist of a partial differential equation (PDE) for the
  microscale temperature along with five ordinary differential equations
  (ODEs) for phase interface locations and root water volume, coupled
  nonlinearly through source terms and algebraic constitutive relations.

\begin{figure}
  \centering
  \begin{tabular}{lll}
    \figlabel{a} & & \figlabel{b} \\[-0.1cm]
    {\setlength{\unitlength}{0.48\textwidth}
      \begin{picture}(1,0.7)
        \put(0,0){\includegraphics[width=0.48\textwidth]{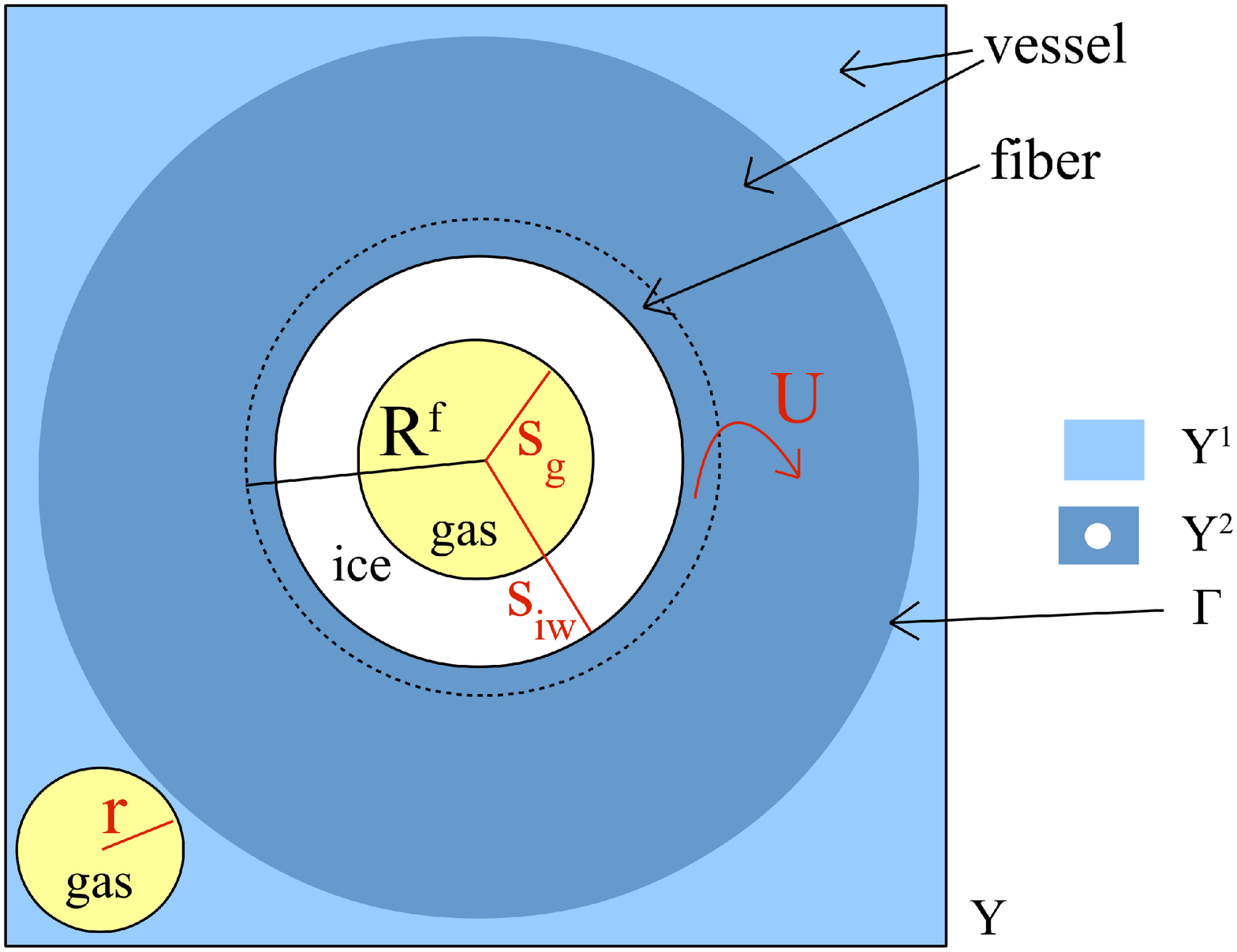}}
          \thicklines
          \linethickness{0.5mm}
          \put(0.39,0.396){\circle*{0.02}}
          \put(0.39,0.39){\vector(0,1){0.26}}
          \put(0.365,0.67){{\bfseries\itshape\large y}}
      \end{picture}
    }
    & & 
    {\setlength{\unitlength}{0.40\textwidth}
      \begin{picture}(1,1)
        \put(0,0){\includegraphics[width=0.40\textwidth]{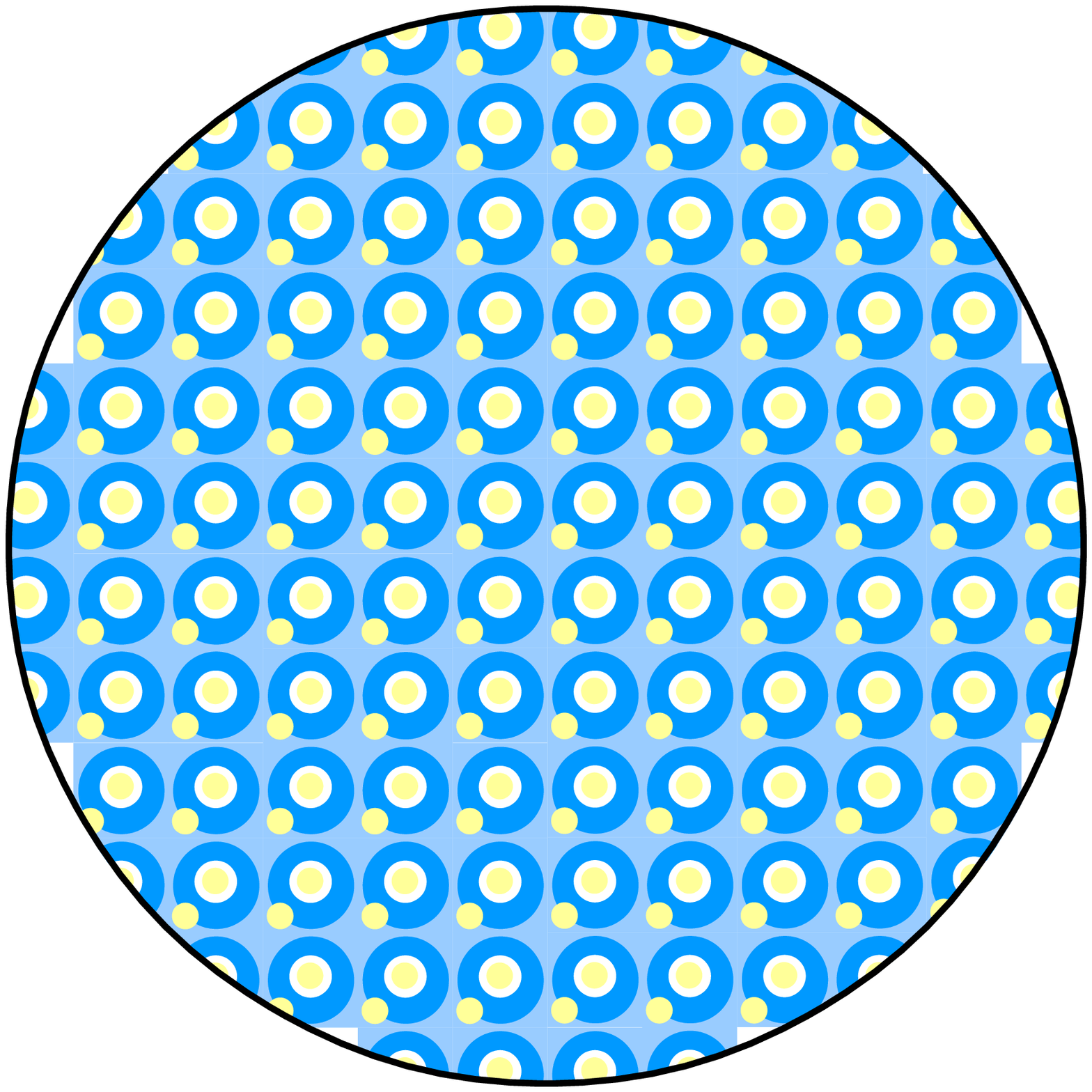}}
        \thicklines
        \linethickness{0.5mm}
        \put(0.5,0.48){\circle*{0.02}}
        \put(0.5,0.48){\vector(0,1){0.28}}
        \put(0.475,0.78){{\bfseries\itshape\Large x}}
        \put(0.85,0.8){{\Large $\mathbf{\Omega}$}}
      \end{picture}
    }
  \end{tabular}
  \caption{{\bf Multiscale problem geometry.}  \figlabel{a} Idealized
    microscale fiber--vessel geometry, consisting of a square reference
    cell $Y$ with side length $\ell$.  \myrevision{This diagram depicts
      a thawing scenario (stage 4 in figure~\ref{fig:cycle-stages})}
    wherein a fiber of radius $R^f$ (dashed line) contains a gas bubble
    surrounded by annular layers of ice and liquid water, where the
    gas/ice and ice/water interfaces are concentric circles of radius
    $s_g$ and $s_{iw}$ respectively.  The vessel contains a gas bubble
    (radius $r$) and liquid sap (water plus sugar). The total liquid
    volume transferred from fiber to vessel is denoted by $U$.  The
    reference cell $Y=Y^1 \cup Y^2$ is divided into two regions
    \myrevision{separated by a curve $\Gamma$}, where diffusion on $Y^1$
    (\myrevision{light blue, outer vessel}) is fast and on $Y^2$
    (\myrevision{dark blue, fiber plus fiber--vessel overlap region}) is
    slow.  \figlabel{b} A requirement for homogenization is that the
    tree cross-section can be approximated by a periodic fine-structured
    domain, tiled with copies of the reference cell.  The macroscale
    problem is then solved on a homogeneous domain $\Omega$ having the
    same size.  \myrevision{Radial coordinates on the micro- and
      macroscales are denoted $y$ and $x$ respectively.}}
  \label{fig:homo}
\end{figure}

\item Supplement the thawing model with analogous equations for the
  freezing process, which have similar structure but differ slightly
  depending on the precise state of freezing or thawing in the
  fibers and vessels.

\item Apply periodic homogenization~\cite{allaire-1992,
    graf-stockie-2014} to derive a macroscopic equation for temperature
  that is coupled to the microscale (reference cell) problem at each
  point within the tree stem (see figure~\ref{fig:homo}b).  The
  macroscopic heat diffusion equation contains an integral source term
  depending on the microscale temperature and capturing all processes
  on the cellular level.  A similar approach has been applied to 
  studying protein-mediated transport of water and solutes in non-woody
  plant tissues~\cite{chavarriakrauser-ptashnyk-2013}.

\item Exploit radial symmetry on the micro- and macroscales to reduce
  both PDEs for temperature to a single spatial dimension.
  \myrevision{We will see later in section~\ref{sec:sims}a that the
    microscale equations need only be solved on the circular sub-region
    $Y^2$ in figure~\ref{fig:homo}a (consisting of the fiber and
    surrounding vessel overlap region) which is clearly radially
    symmetric.}
\end{itemize}

\subsection{Microscale equations for cell-level thawing process}
%
The cell-level model is based on equations already developed for the
thawing half of the freeze--thaw process by Ceseri and
Stockie~\cite{ceseri-stockie-2013}, which were subsequently homogenized
by Graf and Stockie~\cite{graf-stockie-2014}.  We therefore begin by
considering an intermediate state in the thawing process corresponding
to stage~4 in figure~\ref{fig:cycle-stages}, during which the vessel sap
is completely thawed while the fiber contains both liquid and ice.  We
extend the Ceseri--Stockie model by incorporating additional physical
effects that capture the influence of \myrevision{ice--water surface
  tension}, root water uptake, and volume change due to ice/water phase
transitions.  We discuss some of the most important assumptions and
modifications next, leaving the reader to consult the
references~\cite{ceseri-stockie-2013, graf-stockie-2014} for a complete
derivation and discussion of assumptions.

Our model is based on the conceptual diagram in
\myrevision{figure~\ref{fig:sapwood}b} that depicts a single
vessel--fiber pair.  Tracheids are not treated \myrevision{separately}
but instead `lumped together' with vessels because, although they are
connected hydraulically to vessels via paired pits, they have a much
smaller diameter and correspondingly lesser influence on sap transport
than vessels.  Because multiple fibers adjoin and interact hydraulically
with each vessel, we introduce the parameter $N^f$ representing an
average number of fibers per vessel, which is estimated from SEM
images~\cite{cirelli-etal-2008} as $N^f\approx 16$.  Our model captures
the dynamics of a single fiber and then scales all fiber--vessel flux
terms by an appropriate factor of $N^f$.

We assume that sapwood can be represented as a doubly-periodic array of
idealized reference cells $Y$ as pictured in figure~\ref{fig:homo},
where each reference cell contains a circular fiber embedded within a
surrounding square liquid region representing the adjoining vessel.
This choice of geometry is made for mathematical convenience in the
homogenization step, and can be justified because our aim is to derive a
system of equations that captures the net effect of sap flow and heat
transport on the microscale, keeping in mind that any specific geometric
details will ultimately be `averaged out' during the homogenization
process anyways.

\myrevision{Our 2D geometry comes with the built-in assumption that
  axial (vertical) variations are neglected.  In the absence of root
  water uptake, the model tree behaves as a closed system that is
  essentially in equilibrium.  Any pressure differences initiated by
  phase change engender primarily horizontal flow between neighbouring
  cells, and negligible axial flow.  Furthermore, we have already
  shown~\cite{ceseri-stockie-2013} that phase change on the microscale
  dominates the pressure exchange process and occurs very rapidly (on
  the order of milliseconds).  Root water uptake induces an axial flow
  but this is a much slower process; therefore, over the time scales
  that dominate the microscale problem, axial transients may be
  neglected.}

The fiber is a circular cylinder with length $L^f$ and cross-sectional
radius $R^f$ as pictured in figure~\ref{fig:homo}a.  Situated at the
centre of the fiber is a cylindrical gas bubble with time-varying radius
$s_g(t)$, outside of which lies an annular ice layer with outer radius
$s_{iw}(t)$.  The remaining volume extending to the fiber radius $R^f$
contains melt-water from thawed ice.  We note that this configuration is
specific to the thawing process, and the ordering of ice and water
layers would be reversed during freezing.


The vessel is represented by the portion of the reference cell lying
outside the fiber--vessel wall (denoted by a dashed line) and the side
length $\ell$ of the reference cell is chosen so that the vessel
cross-sectional area equals that of a cylinder of radius $R^v$.  Keeping
in mind that there are actually $N^f$ fibers connected to each vessel,
we require that $\ell$ satisfy the area constraint
\begin{linenomath*}
\begin{gather}
  \ell^2=\pi N^f(R^f)^2 + \pi (R^v)^2 .
  \label{eq:l-refcell}
\end{gather}
\end{linenomath*}
Within the vessel is a gas bubble of radius $r(t)$, which is surrounded
by liquid sap owing to the FPD effect that lowers freezing temperature
below that in the fiber.  The cumulative volume of melt-water flowing
through the porous fiber--vessel wall is denoted by $U(t)$ and is
measured positive from fiber to vessel.  The final variable that
determines the local state of the fiber--vessel system is the volume of
root water uptake, denoted $U_{root}(t)$.

We may now formulate a first-order system of five ODEs describing the
time evolution of $s_{iw}$, $s_g$, $r$, $U$ and $U_{root}$.  The fiber
ice--water interface is governed by the Stefan
condition~\cite{alexiades-solomon-1993, crank-1984}
\begin{linenomath*}
\begin{gather}
  \partial_t s_{iw} = - \frac{k_w/\rho_w}{(E_w-E_i)} \, \nabla T_2 \cdot 
  \vec{n} + \frac{\partial_t U}{2\pi s_{iw} L^f},
  \label{eq:micro-swi}
\end{gather}
\end{linenomath*}
where $\nabla T_2 \cdot \vec{n}$ represents the normal temperature
derivative on the interface (i.e., the curve along which temperature
equals the \myrevision{melting (or freezing) point $T_m$}) and the final
term accounts for the volume of water transferred between fiber and
vessel.  \myrevision{This form of the Stefan condition assumes that
  liquid motion induced by phase density differences is
  negligible~\cite{alexiades-solomon-1993}}.  The microscale temperature
$T_2(y,t)$ is obtained as the solution of a heat diffusion equation that
will be stated in the next section, where the microscale spatial
coordinate is $y$.  The parameters $\rho_w$ and $k_w$ denote density and
thermal conductivity of liquid water, while $(E_w-E_i)$ is the enthalpy
difference between water and ice (also called the latent heat or
enthalpy of fusion) at locations where $T_2=T_m$.  \myrevision{The
  effects of thermal expansion are known to be relatively
  small~\cite{wiegand-1906} and so have been neglected here.}

Imposing mass conservation yields an equation for the fiber gas
bubble radius (which in this thawing scenario is a gas--ice
interface)
\begin{linenomath*}
\begin{gather}
  \partial_t s_g =
  - \frac{(\rho_w-\rho_i) s_{iw} \partial_t s_{iw}}{s_g \rho_i} +
  \frac{\rho_w \partial_t U}{2\pi s_g \rho_i L^f},
  \label{eq:micro-sgi}
\end{gather}
\end{linenomath*}
where $\rho_i$ is the density of ice.  An equation for the vessel gas
bubble radius follows from a similar mass conservation argument
\begin{linenomath*}
\begin{gather}
  \partial_t r = - \frac{N^f\partial_t U + 
    \partial_t U_{root}}{2\pi r L^v},
  \label{eq:micro-r}
\end{gather}
\end{linenomath*}
where $L^v$ denotes the length of a vessel.  This last equation
expresses the balance between water flux from neighbouring fibers and
the slight volume change stemming from the water/ice density
difference. \myrevision{The effect of gas dissolution has been omitted
  here but will be incorporated below in the gas density; this
  approximation was already justified in \cite{ceseri-stockie-2014},
  which showed that incorporating dissolution in these equations has
  negligible impact on the bubble radii $s_g$ and $r$.}

Darcy's law governs liquid water flux through the porous fiber--vessel
wall
\begin{linenomath*}
\begin{gather}
  \partial_t U = -\frac{\hydcond \cellarea}{N^f}\, \Big[ p_w^v(t) -
  p_w^f(t) - p_{osm} + p_i^f(t) \Big],
  \label{eq:micro-U} 
\end{gather}
\end{linenomath*}
where the wall is characterized by hydraulic conductivity $\hydcond$ and
surface area $\cellarea$.  The pressure term in square parentheses
derives from four contributions: liquid pressure in the vessel ($p_w^v$)
and fiber ($p_w^f$), osmotic pressure ($p_{osm}$), and capillary
pressure ($p_i^f$) due to ice--water surface
tension~\cite{fowler-krantz-1994}.  This latter contribution,
\myrevision{also known as cryostatic suction,} follows hand-in-hand with
FPD and arises whenever ice lies on the inside surface of the wall and
liquid sap is present on the vessel side, since then the small capillary
pores in the adjoining wall (with radius $r_{cap}$) contain both ice and
liquid.  For the thawing scenario under consideration here, water lies
on both sides of the fiber--vessel wall and so $p_i^f=0$; however, other
stages in the freeze--thaw process can give rise to non-zero $p_i^f$ as
detailed in the next section (see also figure~\ref{tab:phase-eqns}).

The final ODE comes from another application of Darcy's law to root
flux
\begin{linenomath*}
\begin{gather}
  \partial_t U_{root} = \max \Big\{ -\hydcond_r \cellarea_r
  \big( p_w^v(t) - p_{soil} \big), \; 0 \Big\}   ,
  \label{eq:micro-Uroot} 
\end{gather}
\end{linenomath*}
where $\hydcond_r$ is the root hydraulic conductivity and $\cellarea_r$
denotes the portion of root surface area corresponding to a single
vessel.  The cut-off function `$\max\{\,\cdot\,,0\}$' ensures that water
only flows inward from soil to roots and not outward, which is
consistent with \myrevision{experiments that demonstrate root outflow
  can be a factor of five smaller than that for
  inflow~\cite{henzler-etal-1999}.  Indeed, studies of root water
  transport in a variety of tree species show that root conductivity can
  vary with factors such as temperature~\cite{fennell-markhart-1998},
  root age~\cite{dawson-1997}, and time of day \cite{henzler-etal-1999}
  or season \cite{mcelrone-etal-2007}.  Many authors attribute this
  selective control of water transport to membrane proteins known as
  aquaporins~\cite{javot-maurel-2002}.}
%
%
%

\begin{table}
  \footnotesize
  \centering
  \caption{{\bf Model parameters for base case simulation}}
  \longcaption{Unless cited otherwise, all parameter values are
    taken from~\cite{ceseri-stockie-2013}}
  \label{tab:params}
  \footnotesize
  \begin{tabular}{clrlc}\hline
    {\bf Symbol} & {\bf Description} 
    & \mymc{{\bf Values}} & {\bf Units} \\\hline
    & & & & \\[-0.2cm]
    \multicolumn{5}{l}{\emph{Microscale variables (functions of time $t$
        and space $x,y$):}}\\
    $s_{iw}$, $s_g$ & interface locations in fiber & & & m \\ 
    $r$ & vessel bubble radius & & & m \\
    $U$ & water volume flowing from fiber to vessel & & & m${}^3$ \\
    $U_{root}$ & root water volume uptake & & & m${}^3$ \\
    $V$ & volume & & & m${}^3$ \\
    $p$ & pressure & & & Pa \\
    $\rho$ & density & & & kg\3m${}^{-3}$ \\
    & & & & \\[-0.2cm]
    \multicolumn{5}{l}{\emph{Subscripts:}  $i$, $w$, $g$ for ice,
      water/sap, gas}\\
    & & & & \\[-0.2cm]
    \multicolumn{5}{l}{\emph{Superscripts:} $f$, $v$ for fiber, vessel}\\
    & & & & \\[-0.2cm]
    \multicolumn{5}{l}{\emph{Tree structural parameters:}}\\
    $\cellarea$ & area of fiber--vessel wall & 
    \mymc{\myee{6.28}{-8}} & m${}^2$ \\
    $\cellarea_r$& root area for a single vessel~\cite{day-harris-2007} & 
    \mymc{\myee{1.14}{-6}} & m${}^2$ \\
    $\ell$      & side length of reference cell, equation~\eqref{eq:l-refcell} & 
    \mymc{\myee{4.33}{-5}} & m \\
    $L^f$       & length of fiber & 
    \mymc{\myee{1.0}{-3}} & m \\
    $L^v$       & length of vessel \myrevision{element} & 
    \mymc{\myee{5.0}{-4}} & m \\
    $\hydcond$     & conductivity of fiber--vessel wall & 
    \mymc{\myee{5.54}{-13}} & m\3s${}^{-1}$\3Pa${}^{-1}$ \\
    $\hydcond_r$   & conductivity of
    roots~\cite{steudle-peterson-1998, tyree-etal-1994} & 
    \mymc{\myee{2.7}{-16}} & m\3s${}^{-1}$\3Pa${}^{-1}$ \\
    %
    %
    $N^f$       & number of fibers per vessel & 
    \mymc{$16$} & -- \\
    $R^f$       & inside radius of fiber & 
    \mymc{\myee{3.5}{-6}} & m \\
    $R^v$       & inside radius of vessel & 
    \mymc{\myee{2.0}{-5}} & m \\
    $r_{cap}$   & radius of pores in fiber--vessel
    wall~\cite{khaddour-etal-2010} &  
    \mymc{\myee{2.80}{-7}} & m \\
    $W$         & thickness of fiber--vessel wall & 
    \mymc{\myee{3.64}{-6}} & m \\
    & & & & \\[-0.2cm]
    \multicolumn{2}{l}{\emph{Water phase properties:}} & 
    \emph{ice,} & \emph{liquid} & \\
    $c_i$, $c_w$ & specific heat capacity & $2100$, & $4180$ 
    & J\3\degK${}^{-1}$\3kg${}^{-1}$ \\
    $E_i$, $E_w$ & enthalpy at $T_m$ & $574$, & $907$ & kJ\3kg${}^{-1}$ \\
    $k_i$, $k_w$ & thermal conductivity & $2.22$, & $0.556$ 
    & W\3m${}^{-1}$\3\degK${}^{-1}$ \\
    %
    %
    %
    $\rho_i$, $\rho_w$ & density & $917$, & $1000$  
    & kg\3m${}^{-3}$ \\
    $\sigma_{iw}$, $\sigma_{gw}$ & surface
    tension~\cite{fowler-krantz-1994} & $0.033$, & $0.076$ & N\3m${}^{-1}$ \\
    $c_\infty$   & regularization parameter, equation~\eqref{eq:T-E-1} & 
    \mymc{\myee{1.0}{7}} & J\3\degK${}^{-1}$\3kg${}^{-1}$ \\
    & & & & \\[-0.2cm]
    \multicolumn{5}{l}{\emph{Physical constants:}}\\
    $H$         & Henry's constant for air in water & 
    \mymc{$0.0274$} & -- \\
    $K_b$       & cryoscopic (Blagden) constant & 
    \mymc{$1.853$} & kg\3\degK\3mol${}^{-1}$ \\
    $M_g$       & molar mass of gas (air) & 
    \mymc{$0.029$} & kg\3mol${}^{-1}$ \\
    $\mathscr{R}$     & universal gas constant & 
    \mymc{$8.314$} & J\3\degK${}^{-1}$\3mol${}^{-1}$ \\
    $T_m$       & \myrevision{melting point for pure water} & 
    \mymc{$273.150$}& \degK \\
    & & & & \\[-0.2cm]
    \multicolumn{5}{l}{\emph{`Base case' simulation:}}\\
    $C_s$       & sap sugar concentration (3\% by mass) & 
    \mymc{$87.6$} & mol\3m${}^{-3}$ \\
    $p_{soil}$  & soil pressure at roots $= p_w^v(0)$ & 
    \mymc{\myee{2.03}{5}} & Pa \\
    $R$         & tree cross-sectional radius & 
    \mymc{$0.035$} & m \\
    $T_a(t)$    & ambient temperature & 
    \mymc{$[-10,20]+T_m$} & \degK \\
    $T_{m,sap}$    & \myrevision{melting} point for sap $=T_m-K_b C_s/\rho_w$ &
    \mymc{$272.988$} & \degK \\
    \hline
  \end{tabular}
\end{table}

In the preceding discussion we introduced a number of constant
parameters whose values are listed in table~\ref{tab:params}.  The
remaining symbols correspond to intermediate variables whose definitions
we provide next.  First, the density of gas in the fiber and
vessel bubbles depends on initial values of density and volume, modified
to account for dissolved gas according to
\begin{linenomath*}
\begin{align}
  \rho_g^f = \left( \frac{V_g^f(0) + HV_w^f(0)}{V_g^f + HV_w^f}
  \right) \rho_g^f(0), &\quad & \rho_g^v = \left( \frac{V_g^v(0) +
      HV_w^v(0)}{V_g^v + HV_w^v} \right)
  \rho_g^v(0), \label{eq:micro-constit-rhog}
\end{align}
\end{linenomath*}
where $H$ is the dimensionless Henry's constant for air in water. The
various phase volumes are determined from the cylindrical cell geometry
as
\begin{linenomath*}
\begin{alignat}{3}
  V_g^f &= \pi L^f s_g^2, \qquad&  V_g^v &= \pi L^v r^2,
  \label{eq:micro-constit-volume1}\\
  V_w^f &= \pi L^f \left( \left(R^f\right)^2 - s_{iw}^2 \right), 
   \qquad &
  V_w^v &= \pi L^v \left( \left(R^v\right)^2 - r^2 \right).
  \label{eq:micro-constit-volume2}
\end{alignat}
\end{linenomath*}
The corresponding gas pressures are given by the ideal gas law as
\begin{linenomath*}
\begin{align}
  p_g^f = \frac{\rho_g^f \mathscr{R} T}{M_g}, \qquad 
  p_g^v = \frac{\rho_g^v \mathscr{R} T}{M_g}, 
  \label{eq:micro-constit-pg}
\end{align}
\end{linenomath*}
where $\mathscr{R}$ is the universal gas constant and $M_g$ is the molar
mass of air.  The water and gas pressures in both fiber and vessel
differ by an amount equal to the capillary pressure, which is determined
by the Young--Laplace equation as
\begin{linenomath*}
\begin{align}
  p_w^f = p_g^f - \frac{2\sigma_{gw}}{s_g}, \qquad 
  p_w^v = p_g^v - \frac{2\sigma_{gw}}{r},
  \label{eq:micro-constit-pw}
\end{align}
\end{linenomath*}
where $\sigma_{gw}$ is the air--water surface tension.  The osmotic
pressure across the fiber--vessel wall depends on sap sugar
concentration according to
\begin{linenomath*}
\begin{gather}
  p_{osm} = C_s \mathscr{R} T.
  \label{eq:micro-constit-posm}
\end{gather}
\end{linenomath*}
Finally, the sap sugar content induces a reduction in freezing
temperature that obeys
\begin{linenomath*}
\begin{gather}
  T_{m,sap} = T_m - \Delta T_{fpd} = T_m - \frac{K_b C_s}{\rho_w}.
  \label{eq:micro-constit-Tcsap}
\end{gather}
\end{linenomath*}

\subsection{Equations for other phase transitions}
%
In the previous section we developed equations specific to the thawing
process, during which the vessel is completely thawed and the fiber
contains a mix of gas, water and ice (see stage~4 in
figure~\ref{fig:cycle-stages}).  We describe next how these equations
should be modified to capture other freeze--thaw states in the fiber and
vessel.  In particular, we account for the fact that phase interfaces
can appear or disappear whenever ice completely thaws (or liquid
completely freezes), as well as the reversal of the ice and water layers
in the fiber during freezing and thawing.

\begin{figure}
  \centering
  \footnotesize
  \renewcommand{\arraystretch}{1.3}
  \begin{tabular}{l@{}l@{}l}\hline
    \parbox[t]{2.1cm}{\centering 1 \\Completely \\ thawed} & 
    \raisebox{-1.7cm}{\includegraphics[width=4cm]{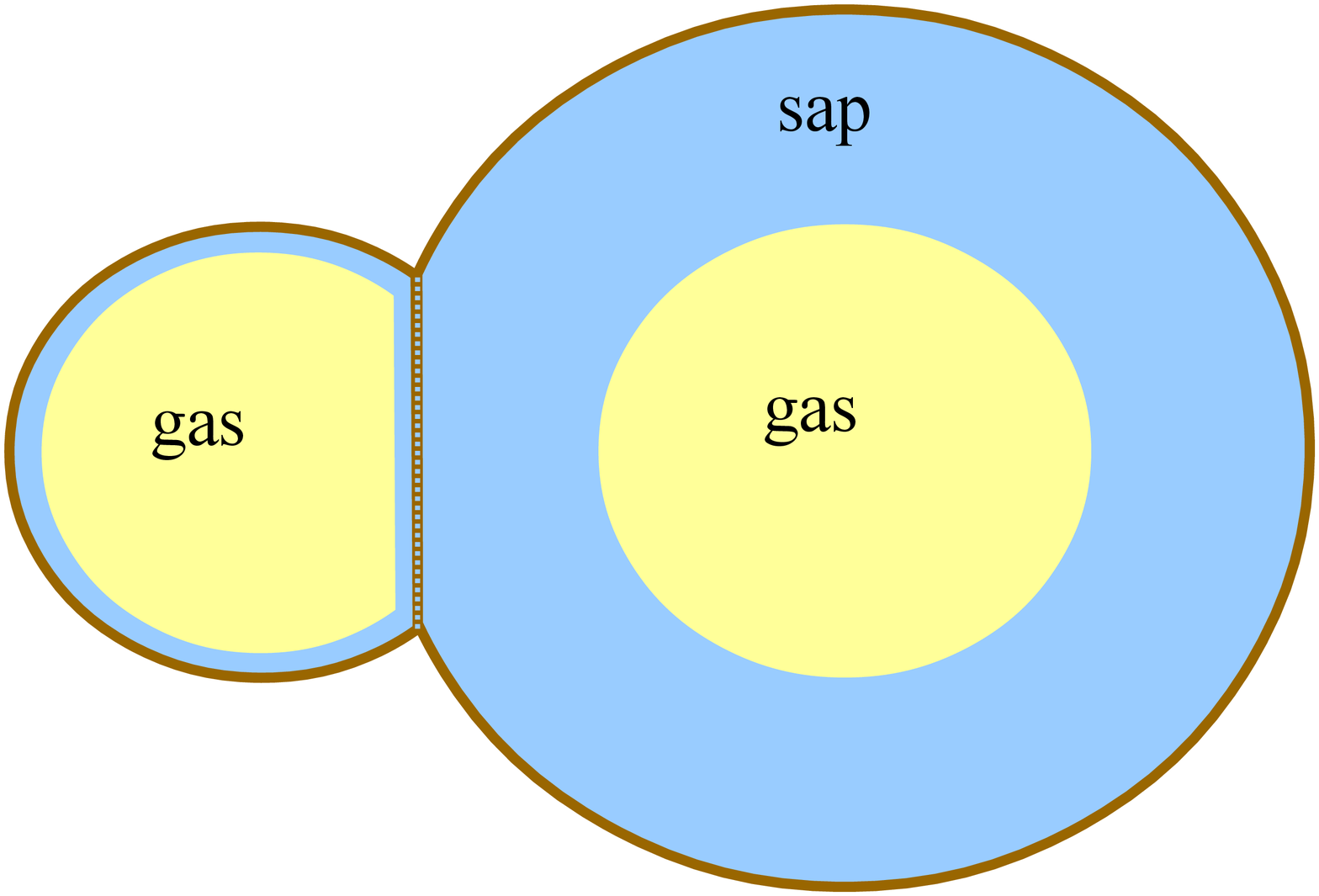}} & 
    \begin{minipage}{7.8cm}
        \begin{eqnarray}
          p_i^f             &=& 0\\
          \partial_t s_{iw} &=& 0\\
          \partial_t s_g    &=& \frac{\partial_t U}{2\pi s_g L^f}
        \end{eqnarray}
    \end{minipage}
    \\\hline
    \parbox[t]{2.6cm}{\centering 2 \\Vessel thawed \\ Fiber freezing} & 
    \raisebox{-2cm}{\includegraphics[width=4cm]{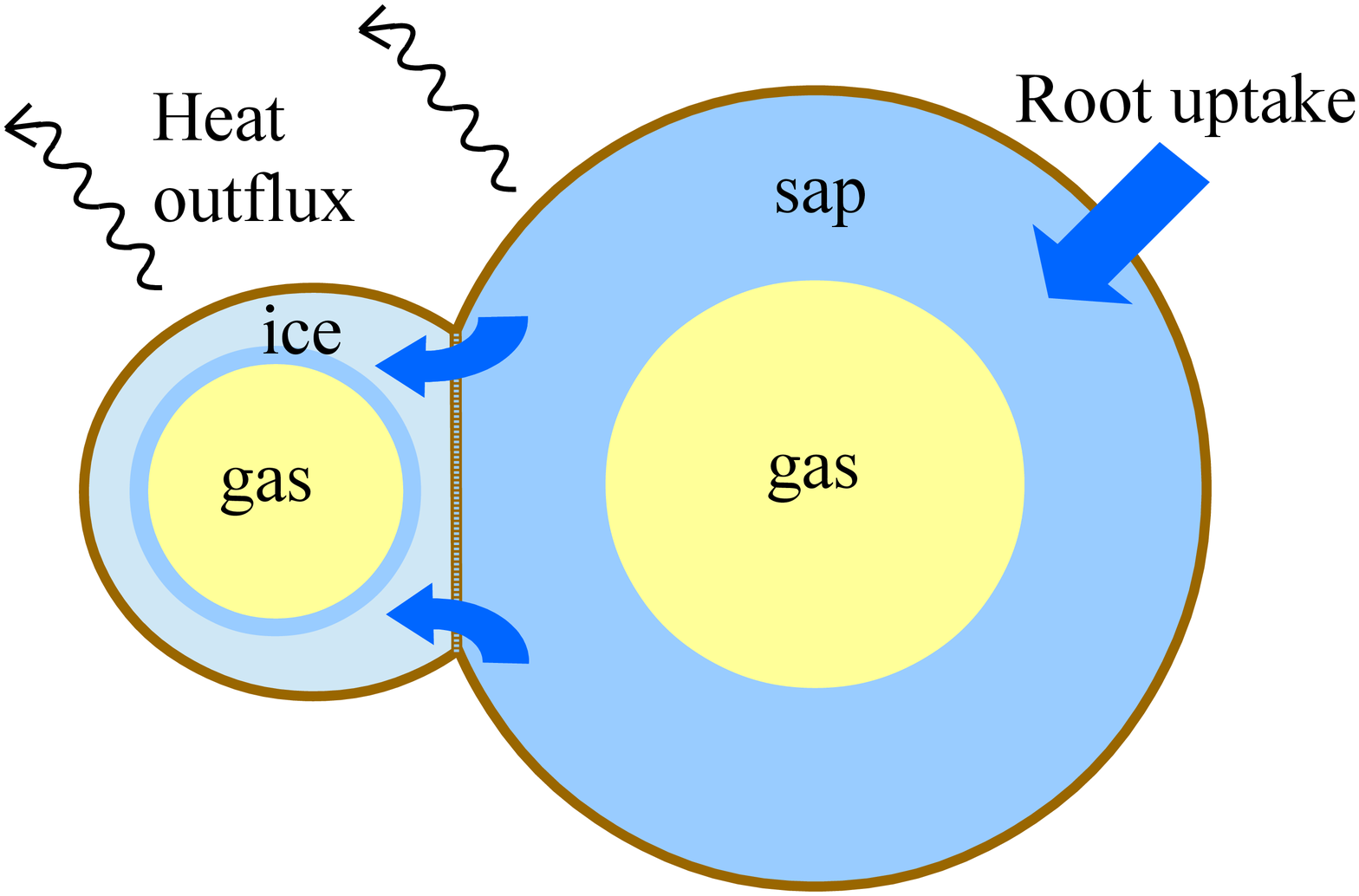}} & 
    \begin{minipage}{7.8cm}
        \begin{eqnarray}
          p_i^f &=& \frac{2\sigma_{iw}}{r_{cap}} \frac{V_i^f}{V_i^f + V_w^f}\\
          \partial_t s_{iw} &=& \frac{k_i/\rho_i}{(E_w - E_i)} 
          \nabla T\cdot\vec{n} + \frac{\partial_t 
            U \rho_w}{2\pi s_{iw} L^f \rho_i}\\
          \partial_t s_g &=& \frac{(\rho_w-\rho_i) s_{iw} \partial_t
            s_{iw}}{s_g \rho_w} + \frac{\partial_t U}{2\pi 
            s_g L^f} 
        \end{eqnarray}
    \end{minipage}
    \\\hline
    \parbox[t]{2.6cm}{\centering Vessel freezing \\ Fiber frozen} & 
		\raisebox{-1.5cm}{\includegraphics[width=4cm]{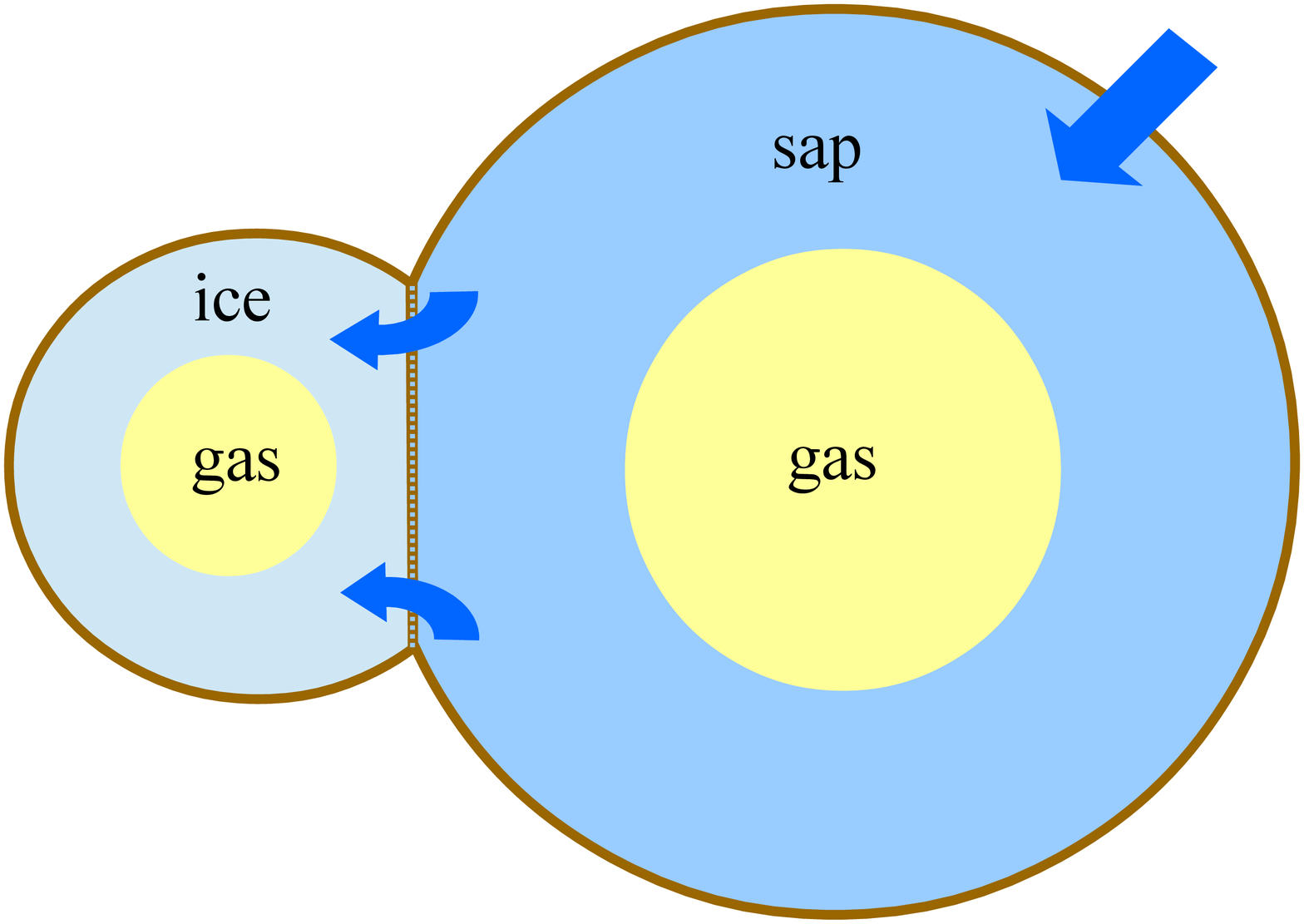}} & 
    \begin{minipage}{7.8cm}
        \begin{eqnarray}
          p_i^f &=& \frac{2\sigma_{iw}}{r_{cap}}\\
          \partial_t s_{iw} &=& 0\\
          \partial_t s_g &=& \min\left\{ \frac{\rho_w \partial_t U}{2\pi
              s_g L^f \rho_i}, \; 0\right\}
        \end{eqnarray}
    \end{minipage}
    \\\hline
    \parbox[tc]{2.6cm}{\centering 3 \\Completely \\ frozen} & 
    \raisebox{-1.5cm}{\includegraphics[width=4cm]{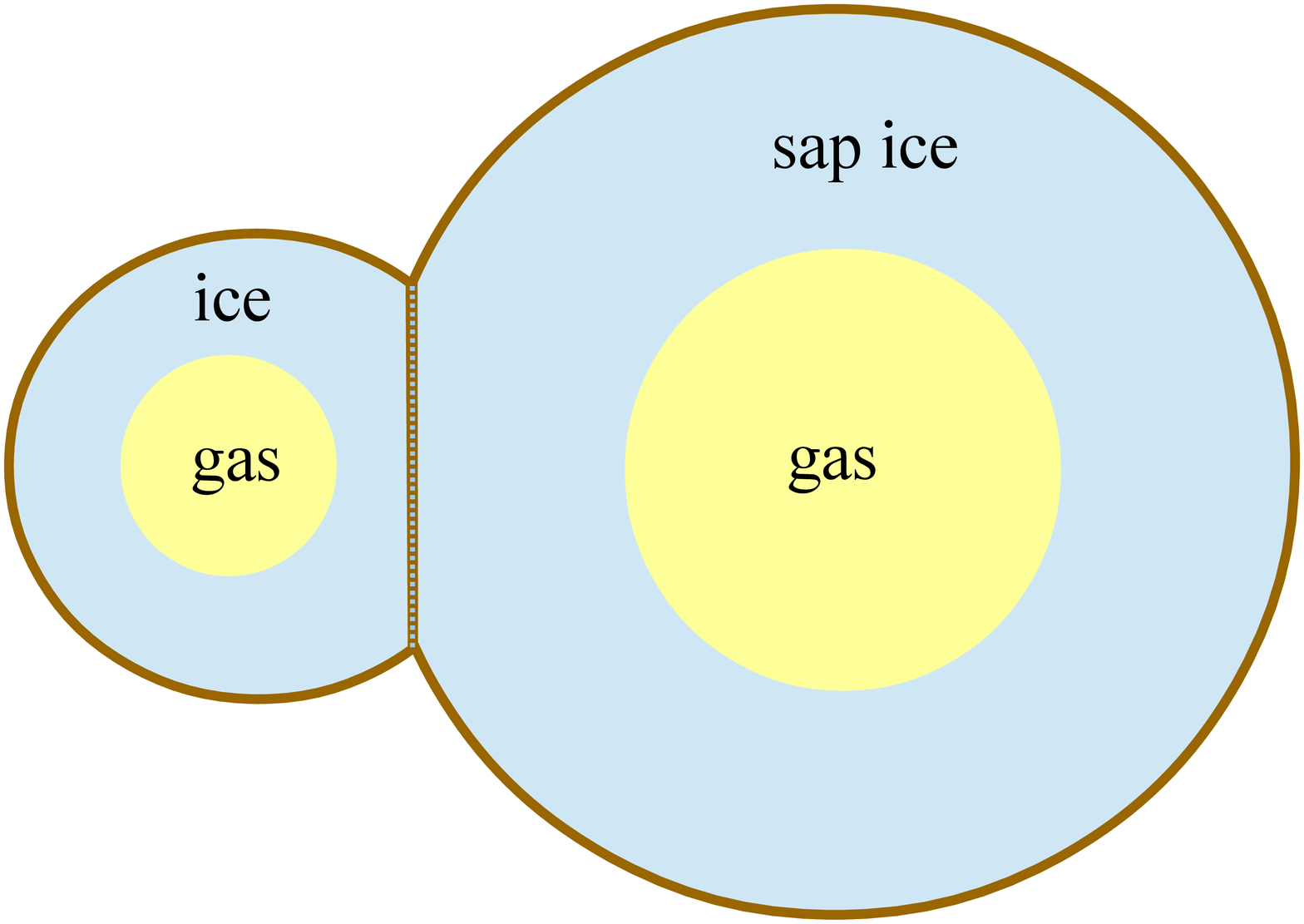}} & 
    \begin{minipage}{7.8cm}
        \begin{eqnarray}
          p_i^f = 0  &\quad& \partial_t r = 0 \\
          \partial_t s_{iw} = 0 &\quad& \partial_t U = 0\\
          \partial_t s_g = 0 &\quad& \partial_t U_{root} = 0
        \end{eqnarray}
    \end{minipage}
    \\\hline
    \parbox[t]{2.6cm}{\centering Vessel thawing \\ Fiber frozen} & 
		\raisebox{-1.5cm}{\includegraphics[width=4cm]{frozen-liquid}} & 
    \begin{minipage}{7.8cm}
        \begin{eqnarray}
          p_i^f &=& \frac{2\sigma_{iw}}{r_{cap}}\\
          \partial_t s_{iw} &=& 0\\
          \partial_t s_g &=& \min\left\{ \frac{\rho_w \partial_t U}{2\pi
              s_g L^f \rho_i}, \; 0\right\}
        \end{eqnarray}
    \end{minipage}
    \\\hline
    \parbox[t]{2.6cm}{\centering 4 \\Vessel thawed \\ Fiber thawing} &
    \raisebox{-1.8cm}{\includegraphics[width=4cm]{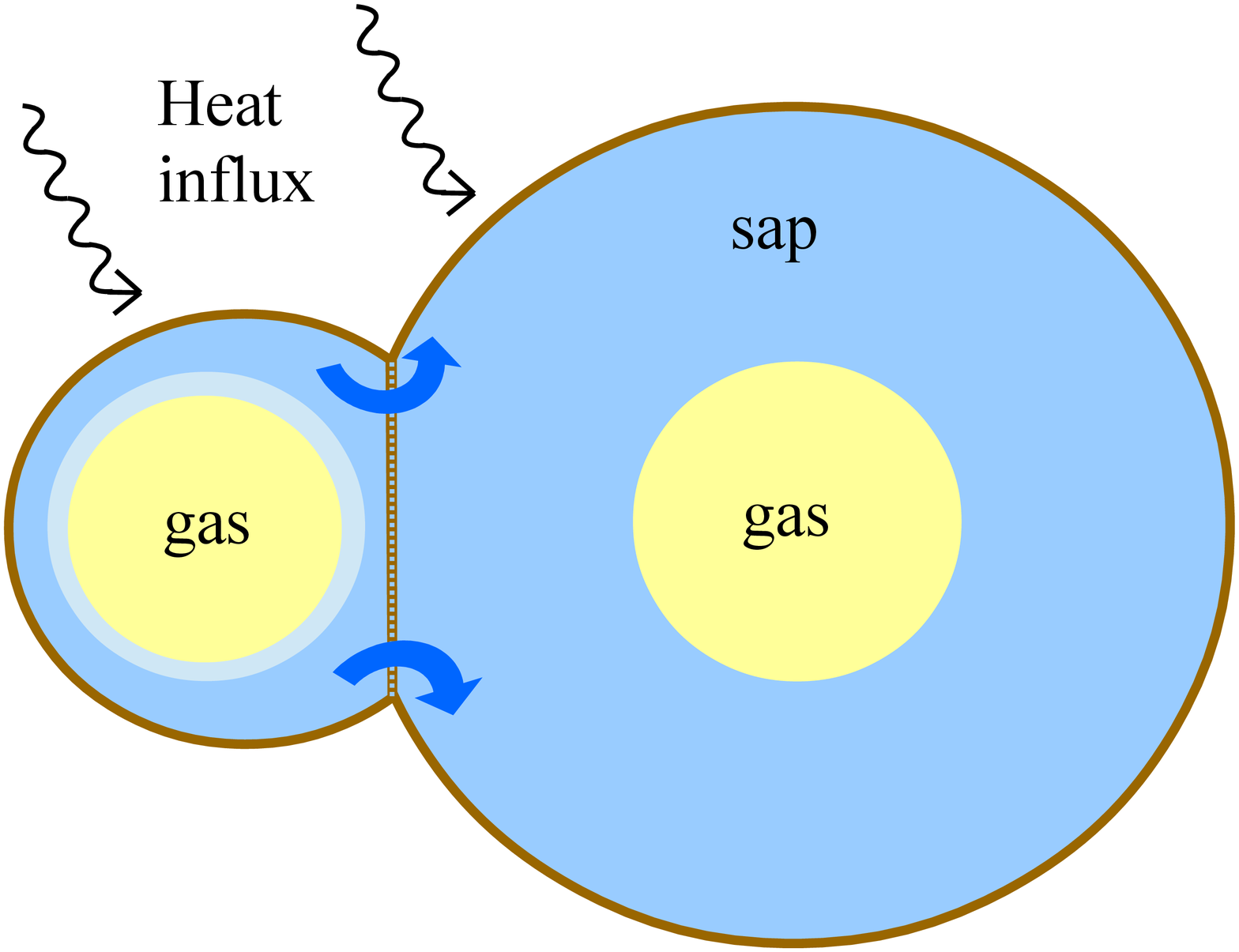}} & 
    \begin{minipage}{7.8cm}
        \begin{eqnarray}
          p_i^f &=& 0\\
          \partial_t s_{iw} &=& -\frac{k_w/\rho_w}{(E_w - E_i)} 
          \nabla T\cdot\vec{n} + \frac{\partial_t U}{2\pi s_{iw} L^f}\\ 
          \partial_t s_g &=& - \frac{(\rho_w-\rho_i) s_{iw} \partial_t
            s_{iw}}{s_g \rho_i} + \frac{\rho_w \partial_t U}{2\pi s_g 
            L^f \rho_i} 
        \end{eqnarray}
    \end{minipage}
    \\\hline
  \end{tabular}
  \caption{{\bf Microscale equations for all stages of the freeze--thaw 
    process.}}
  \label{tab:phase-eqns}
\end{figure}

In fact, many equations remain unchanged throughout the entire
freeze--thaw cycle, with the exception being those for $s_g$, $s_{iw}$
and $p_i^f$.  The required modifications for each case are listed in
figure~\ref{tab:phase-eqns}, referenced by the numbered stages in
figure~\ref{fig:cycle-stages}.  We emphasize that
\myrevision{the ice--water interface lies within pores in the
  fiber--vessel wall and forms a \emph{mushy layer} wherein both solid
  and liquid phases coexist in the pore space.  This type of phase
  interface (called a \emph{frozen fringe} in the context of ice lensing
  in soils~\cite{fowler-krantz-1994}) differs from an idealized
  gas--water interface in that the interfacial pressure jump $p_i^f$
  increases with ice volume fraction according
  to} 
\begin{linenomath*}
\begin{gather*}
  p_i^f = \frac{2\sigma_{iw}}{r_{cap}}\ \frac{V_i^f}{V_i^f+V_w^f},
\end{gather*}
\end{linenomath*}
where $\sigma_{iw}$ represents the ice--water surface tension and
$V_{i,w}^f$ are the corresponding volume fractions.  Finally, we note
that when both fiber and vessel are completely frozen (stage~3) the
equations for $r$, $U$ and $U_{root}$ also drop out of the system.

\subsection{Homogenized equation for temperature}
%
The equations derived in the preceding two sections govern microscale
processes at the cellular level whereas on the macroscale the
temperature is of primary interest, and it is transport of heat between
the external (ambient) environment and the interior of the tree stem
that drives the freeze--thaw process.  Clearly, there exists a two-way
interaction between the global temperature and the local fiber--vessel
state, wherein temperature governs phase change dynamics in fibers and
vessels, while cellular processes in turn influence heat transport
through the Stefan condition and local phase volume fractions.  To
simplify this complex multiscale problem, we exploit a separation in
spatial scales reflected in the fact that state variables describing the
fiber--vessel configuration are essentially `invisible' on the
macroscale except through their effect on heat transport properties of
the sap- and gas-filled wood.

Because of the repeating microstructure of wood, this problem is ideally
suited to the application of \emph{periodic homogenization}.  The
philosophy behind this approach is to solve at each point in space a
local problem on a reference cell $Y$ that determines the solution state
on the microscale.  By using an appropriate homogenization or averaging
procedure, the effect of microscale variables on the macroscale may then
be incorporated into equations for the global solution variables.  One
technical requirement is that the reference cell must divide into two
sub-regions, $Y=Y^1 \cup Y^2$, separated according to whether heat
diffusion is fast (\myrevision{in $Y^1$, the outer portion of} the
vessel) or slow (\myrevision{in $Y^2$, an overlap region covering the
  fiber and the remainder of the vessel}).  The result is two heat
equations: one governing temperature on the macroscopic domain $\Omega$
and the second on $Y^2\times\Omega$.  When these two equations are
coupled together, we obtain a two-scale temperature solution on the
domain $Y\times\Omega$.  Instead of fully coupling the micro- and
macroscale equations, this homogenization approach leads naturally to a
simpler system of equations that captures the essential aspects of
coupling between scales.  A similar homogenization approach has been
applied by Chavarr\'{\i}a-Krauser and Ptashnyk to a model of water and
solute transport in plants~\cite{chavarriakrauser-ptashnyk-2013}.
%

The dynamics of heat transport are best described using a mixed
formulation written in terms of both temperature and specific enthalpy,
which are denoted respectively by $T_2(x,y,t)$ and $E_2(x,y,t)$ on the
reference cell region $Y^2$, and $T_1(x,t)$ and $E_1(x,t)$ on the
macroscale.  The variables $T_1$ and $E_1$ depend on time $t$ and the
global spatial coordinate $x$, whereas microscale quantities have an
additional dependence on the reference cell $Y$ through a local spatial
coordinate $y$.  Temperature and enthalpy are not independent variables
but instead are related via the piecewise linear function
\begin{linenomath*}
\begin{gather}
  T(E) = 
  \left\{\begin{array}{cl}
      \frac{1}{c_i} E,                 & \mbox{if $E < E_i-\delta_i$},\\[0.2cm]
      T_m+\frac{2E-E_i-E_w}{2c_\infty},& \mbox{if $E_i-\delta_i \leq E < E_w+\delta_w$},\\[0.2cm]
      T_m+\frac{1}{c_w}(E-E_w),        & \mbox{if $E_w+\delta_w \leq E$}.
    \end{array}\right.
  \label{eq:T-E-1}
\end{gather}
\end{linenomath*}
%
We introduce the large parameter $c_\infty$ (taking $c_\infty=10^{7}$ in
practice) to impose a small but nonzero slope $({1}/{c_\infty})$ on
the central plateau region where $T\approx T_m$.  We also make use of
the fact that $E_i=c_i T_m$ and choose
\begin{linenomath*}
\begin{gather}
   \delta_i = \frac{c_i(E_w-E_i)}{2(c_\infty-c_i)} 
   \quad \mbox{and} \quad 
   \delta_w = \frac{c_w(E_w-E_i)}{2(c_\infty-c_w)}, 
   \label{eq:T-E-2}
\end{gather}
\end{linenomath*}
so that the function $T(E)$ is continuous.  This form of $T(E)$ is a
regularization of the exact temperature--enthalpy
relationship~\cite{visintin-1996} that avoids numerical instabilities in
the calculation of temperature and also recovers the exact (piecewise
linear) result in the limit as $c_\infty \to \infty$ and
$\delta_i,\delta_w\to 0$.


During the homogenization procedure~\cite{graf-stockie-2014}, we find
that heat transport in the reference cell must only be
treated on the sub-region $Y^2$ where temperature obeys
\begin{linenomath*}
\begin{gather}
  \label{eq:freezing-T2}
  c_w\partial_t T_2 - \nabla_y\cdot(D(\myrevision{E_2})\nabla_y T_2) = 0 
  \qquad \mbox{in $Y^2(x,t)\times\Omega$},
\end{gather}
\end{linenomath*}
and $D(E_2)$ is a thermal diffusion coefficient that is a piecewise
linear and continuous function of enthalpy~\cite{visintin-1996}
\begin{linenomath*}
\begin{gather}
  D(E) = \left\{\begin{array}{cl}
      \frac{k_i}{\rho_i}, & \mbox{if } E<E_i,\\
      \frac{k_i}{\rho_i} + \frac{E-E_i}{E_w-E_i}
      \left(\frac{k_w}{\rho_w} - \frac{k_i}{\rho_i}\right), 
      & \mbox{if } E_i\leq E < E_w, \\ 
      \frac{k_w}{\rho_w}, & \mbox{if } E_w \leq E.
    \end{array}\right.
\end{gather}
\end{linenomath*}
\myrevision{We employ this nonstandard definition of $D$ (instead of the
  usual thermal diffusivity having units m${}^2$\3s${}^{-1}$) so that we
  can factor out the specific heat, thereby allowing the same
  coefficient to be used in both this microscale heat equation and the
  mixed temperature--enthalpy form we develop below for the macroscale.}
We include an explicit time- and global space-dependence in $Y^2(x,t)$
to emphasize the fact that the ice region within the fiber is bounded by
a moving water--ice interface, and that the fiber configuration varies
from point to point throughout the tree stem.  On the water-ice
interface (corresponding to the inner boundary of $Y^2$) the temperature
equals the \myrevision{melting point} value
\begin{linenomath*}
\begin{gather}
  \label{eq:freezing-T2-bc}
  T_2 = T_m     \qquad \mbox{on $\partial Y^2(x,t)\times\Omega$}.
\end{gather}
\end{linenomath*}
We thereby obtain the macroscale temperature equation
\begin{linenomath*}
\begin{gather}
  \label{eq:freezing-T1}
  \partial_t E_1 - \nabla_x\cdot (\Pi D(E_1)\nabla_x
  T_1) = \frac{1}{|Y^1|} \int_\Gamma D(E_2)\nabla_y T_2\cdot \vec{n}
  \;\rd S
  \qquad \mbox{in $\Omega$},
\end{gather}
\end{linenomath*}
where the coupling with microscale variables is embodied in a surface
integral term.  The factor $\Pi$ multiplying the diffusion coefficient
is a $2\times 2$ matrix whose entries depend on the reference cell
geometry according to
\begin{linenomath*}
\begin{gather}
  \label{eq:freezing-Pi}
  \Pi_{ij} = \frac{1}{|Y^1|} \int_{Y^1} \left(\delta_{ij} + \nabla_y
    \mu_i \right) \, \rd y,  
\end{gather}
\end{linenomath*}
for $i,j=1,2$.  Here, $\delta_{ij}$ is the Kronecker delta symbol and
$\mu_i(y)$ are solutions of a standard reference cell problem on
$Y^1$~\cite{allaire-1992}.
The temperature on the outer surface of the tree is held at the 
ambient value
\begin{linenomath*}
\begin{gather}
  \label{eq:freezing-T1-bc}
  T_1 = T_a(t) \qquad \mbox{on}\ \partial\Omega.
\end{gather}
\end{linenomath*}
Finally, the micro- and macroscale solutions are coupled by matching
temperature on the interior boundary
\begin{linenomath*}
\begin{gather}
  \label{eq:freezing-T1-T2-bc}
  T_2 = T_1 \qquad \mbox{on $\Gamma\times\Omega$}. 
\end{gather}
\end{linenomath*}

In summary, the governing equations consist of a system of
differential--algebraic equations
\eqref{eq:micro-swi}--\eqref{eq:micro-constit-Tcsap} and
\eqref{eq:freezing-T2}--\eqref{eq:freezing-T2-bc} for the microscale
temperature and fiber--vessel state variables within each local
reference cell.  These are supplemented by equations
\eqref{eq:freezing-T1}--\eqref{eq:freezing-T1-T2-bc} for the macroscale
temperature on $\Omega$.  Both problems are solved at each spatial point
$x\in\Omega$ and the two solutions are coupled by means of the integral
source term in \eqref{eq:freezing-T1} and the boundary condition
\eqref{eq:freezing-T1-T2-bc}.  The geometry of the local reference cell
is also incorporated into the macroscale problem via the (constant)
pre-factors $\Pi$ multiplying the diffusion coefficient in
\eqref{eq:freezing-T1}.

\section{Simulating daily freeze--thaw cycles}
\label{sec:sims}

\subsection{Numerical solution algorithm}

The radial symmetry of both micro- and macroscale domains implies that
all solution variables can be written as functions of a single radial
coordinate and time.  We use a \emph{method of lines} approach and
discretize the temperature variables in space using finite elements,
yielding a large system of time-dependent ODEs.  When combined with the
ODEs and algebraic equations governing microscale fiber--vessel
dynamics, the resulting coupled system is integrated in time using a
standard ODE solver.  The spatial discretization on the two scales
proceeds as follows:
\begin{itemize}
\item \emph{Microscale (cell-level) equations:} The fiber ice
  temperature is assumed to be a uniform 0\3\degC, and gas temperature
  is also taken constant since the thermal diffusivity of gas is
  so much larger than that for either ice or water.  Therefore, the PDE
  \eqref{eq:freezing-T2} for temperature on $Y^2$ must only be solved on
  the annular region between $\Gamma$ and the phase interface $s_{iw}$
  (see figure~\ref{fig:homo}a).  We find that sufficient accuracy is
  obtained for $T_2$ by using only 4 radial grid points within the
  annulus.
  Because the phase interface evolves in time, we use a moving mesh
  approach wherein the motion of grid points introduces an additional
  `grid advection' term that is proportional to the mesh point
  velocity~\cite{huang-russell-2011}.

\item \emph{Macroscale (tree-level) equation:} The tree stem is
  similarly divided into equally-spaced radial points,
  and here we find that taking 20 grid points yields sufficient accuracy
  in $T_1$.  Owing to radial symmetry, the integral source term in
  \eqref{eq:freezing-T1} reduces to multiplication by the curve length
  $|\Gamma|$.  The factors $\Pi$ depend only on the reference cell
  geometry and so can be pre-computed at the beginning of a simulation.
\end{itemize}
We employ an efficient split-step approach where in each time step the
reference cell problem is solved for the microscale variables, and then
the macroscale temperature equation is solved by holding the microscale
variables constant.

The algorithm described above has been implemented in Matlab using the
built-in stiff ODE solver {\tt ode15s} to integrate the equations in
time.  The only algorithmic detail remaining to be described is the
switching between equations required as phase interfaces appear or
disappear.  We can capture this switching simply and robustly using the
{\tt Events} option provided in the ODE solver suite, which signals an
event based on zero-crossings of an `indicator function'.  During any
portion of the freeze--thaw cycle, the indicator function is set equal
to either the thickness of a phase interface or the difference between
the phase temperature and the \myrevision{melting} temperature.  When
the indicator crosses zero, the time integration halts, equations
are modified appropriately, and the ODE solver is restarted using the
new set of equations and taking the current solution as the new initial
state.  The time integration then proceeds until the next phase change
event is signalled.
A typical simulation covering 4 daily temperature cycles requires
between 30--45 minutes of clock time on an Apple MacBook Pro
with~2.3\3GHz quad-core Intel~i7 processor.


\subsection{Choice of parameters} 
%
The algorithm just described is used to simulate freeze--thaw dynamics
in a typical \emph{base case} scenario for which all parameters are
listed in table~\ref{tab:params}.  We take a `sapling'
of diameter 0.07\3m consisting entirely of sapwood.  The sugar content of
maple sap ranges from 1--5\% by mass~\cite{tyree-1983} and so we choose
a representative value of 3\%\ that induces a vessel FPD of $\Delta
T_{fpd}=0.162$\3\degC.
To mimic temperature variations during late winter, we let ambient
temperature vary sinusoidally between $-10$ and $+20$\3\degC\ over a
24-hour period (this range is somewhat extreme but is chosen to
correspond with the experiments of Am\'eglio
\etal~\cite{ameglio-etal-2001} that we will describe shortly).  We begin
with a freezing event and initialize the tree in a thawed state with
uniform temperature $0.35$\3\degC, just slightly above the freezing
point.  Each fiber initially contains gas and water with 75\%\ gas by
volume, whereas the vessel has a much smaller initial gas content of
8\%.

There remain two parameters whose values we have not been able to obtain
reasonable estimates from the literature -- root hydraulic conductivity
$\hydcond_r$ and capillary pore radius $r_{cap}$ -- and so we have had to
adjust their values in order to match numerical results with
experimental data.  First, we choose $\hydcond_r=2.7\times
10^{-16}$\3m\3s${}^{-1}$\3Pa${}^{-1}$ so that pressure and root uptake
vary over time scales similar to those observed in
experiments~\cite{ameglio-etal-2001, tyree-1983}.
%
%
%
Then, we take $r_{cap}=2.8\times 10^{-7}$\3m so that the
exudation pressure build-up is within the observed range of 80 to
150\3kPa~\cite{cirelli-etal-2008, cortes-sinclair-1985}.\
This pore size is also consistent with that measured in other membranes
that hinder transport of sucrose
molecules~\cite{khaddour-etal-2010}.

\begin{figure}
  \centering
  \begin{tabular}{lll}
    \figlabel{a} & & \figlabel{b} \\
    \includegraphics[width=0.42\textwidth]{BC_Tyree3} & &
    \raisebox{0.2cm}{\includegraphics[width=0.50\textwidth]{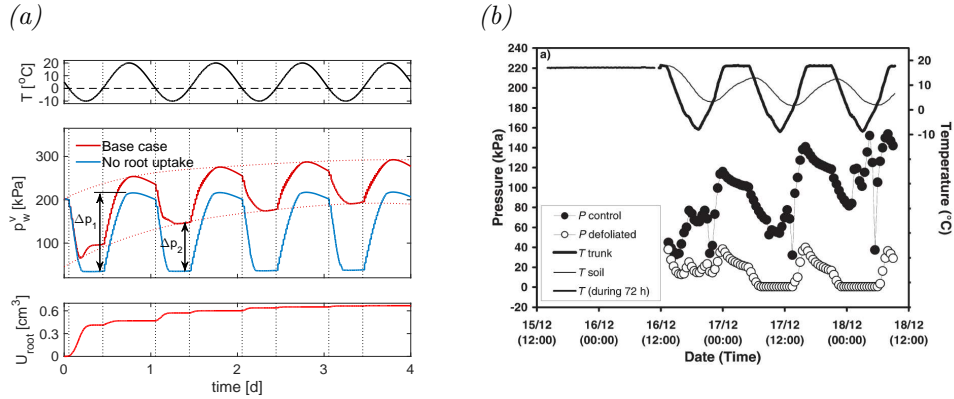}}
  \end{tabular}
%
  \caption{{\bf Comparison of base case simulation with Am\'eglio's
      experiments.} \figlabel{a} Simulated vessel sap pressure (middle,
    with and without root water) and cumulative root water uptake
    (bottom) in response to an imposed periodic ambient temperature
    (top).  The vertical dotted lines highlight times when temperature
    crosses the freezing point.  \myrevision{Two primary features of the
      vessel pressure curve are the amplitude of pressure oscillations
      in each daily cycle ($\Delta p_1$, arising from ice--water
      capillary effects) and the residual pressure increase at the end
      of a cycle ($\Delta p_2$, due to root water uptake).}  %
    \figlabel{b} Am\'eglio \etal's experiments on black
    walnut~\cite{ameglio-etal-2001} (reproduced with permission of
    Oxford University Press). 
    The measurements relevant to our study are
    `P~control' (sap pressure) and `T~trunk' (temperature).
  }
  \label{fig:base}
\end{figure}

\subsection{Base case: Pressure build-up during temperature cycling}
Using these base case parameters and initial conditions, we perform two
numerical simulations: one with root water uptake corresponding to a
soil pressure of $p_{soil}=203$\3kPa, and a second with no root uptake
(e.g., consistent with a completely frozen soil).  Vessel sap pressures
are compared in figure~\ref{fig:base}a, and in both cases we observe a
periodic variation in pressure synchronized with daily temperature
fluctuations.  Without root uptake, the vessel pressure simply
oscillates between two fixed values of 20 and 200\3kPa and there is no
pressure build-up over multiple freeze--thaw cycles.  However, when root
uptake is included there is a gradual pressure increase superimposed on
the background oscillations, with a total increase (measured from the
local maximum in each cycle) of roughly 80\3kPa over the four days. The
accompanying plot of total root uptake in figure~\ref{fig:base}a shows
that the majority of root water is absorbed during the first
freeze--thaw cycle, followed by a more gradual uptake that is
essentially complete after 3 days.

We next draw a direct comparison with the experiments of Am\'eglio
\etal~\cite{ameglio-etal-2001} who studied black walnut trees
(\emph{Juglans nigra}) in a controlled laboratory setting where the
living stump of an excised tree branch was connected via a sealed pipe
to a pressure transducer.  We calculate vessel sap pressure in our
simulations as an average pressure across the stem cross-section to be
as close as possible to such a transducer measurement.  We are unaware
of any comparable data for sugar maple, but we claim that a meaningful
comparison may still be drawn with Am\'eglio's results since black
walnut is closely related to maple and undergoes exudation under similar
conditions~\myrevision{\cite{ameglio-etal-2001, cirelli-etal-2008,
    ewers-etal-2001}}.
The curves to focus on in Am\'eglio's figure~\ref{fig:base}b are the
air temperature (labelled `$T$ trunk') and stem pressure (labelled `$P$
control').

The qualitative agreement between simulated and experimental pressures
is remarkable considering the complexity of the processes involved and
the minimal parameter fitting required.  The overall shape of pressure
curves is similar, with each freeze--thaw cycle exhibiting a rapid
increase whenever temperature exceeds the freezing point.  The pressure
then attains a maximum, after which there is a slight decrease over
roughly 6--8 hours, followed by a rapid drop as ambient temperature
crosses the freezing point again.  We remark that there is also a rough
quantitative agreement between simulations and experiments
\myrevision{in that pressure oscillations have an amplitude of 80 to
  100\3kPa, and the total pressure build-up over four days also is
  similar.}  On the other hand, the maximum value of our simulated
pressure is 290\3kPa, which is almost double the 160\3kPa observed
in the black walnut experiments; however, it is possible that more time
is needed for the experiment to reach steady state, not to
mention that there are species--specific differences that could
influence pressure.

\myrevision{A more quantitative comparison can be drawn based on two
  characteristic features of the pressure in figure~\ref{fig:base}a
  labelled as $\Delta p_1$ and $\Delta p_2$.  The first corresponds to
  the amplitude of oscillations in the absence of root uptake, which
  derives mainly from cryostatic suction and so can be estimated using
  the formula $\Delta p_1 \approx 2\sigma_{iw}/r_{cap} \approx 236$~kPa.
  This value is close to the computed amplitude of the `no root uptake'
  curve, as well as to the rise in vessel pressure during the initial
  thawing event for the `base case'.  The second feature $\Delta p_2$
  captures the exudation pressure build-up during the first freeze--thaw
  cycle which arises mainly from root water uptake.  Because this
  additional water acts to compress the gas in fiber and vessel, we
  apply the differential form of the ideal gas law at constant
  temperature, $\Delta p_2 \approx - p\,{\Delta V}/{V}$, during the
  first freezing event.  Substituting values of $p\approx 200$~kPa for
  the initial vessel pressure, $\Delta V\approx 0.4$~cm$^3$ for the root
  water volume uptake (taken from figure~\ref{fig:base}a) and $V\approx
  1.15$~cm$^3$ for the initial gas volume in a slice through the tree
  cross-section (with thickness equal to that of the reference cell,
  $L^f$), we obtain $|\Delta p_2|\approx 69$~kPa.  The correspondence
  between this estimate and the computed value of 50~kPa is reasonable,
  considering that it ignores effects such as gas dissolution.}

Despite the abundance of experimental data available
for sugar maple~\cite{cirelli-etal-2008, cortes-sinclair-1985,
  johnson-tyree-dixon-1987, tyree-1983},
most experiments measure sap outflux from tapped
trees~\cite{cortes-sinclair-1985} rather than the `closed system'
corresponding to an untapped tree that we consider here.  Other
measurements have been taken of excised wood samples rather than living
trees, while yet others were taken in uncontrolled external conditions
with irregular variations in pressure and ambient temperature.
Consequently, we hesitate to attempt a detailed comparison between any
of these experiments and our simulations; nonetheless, we can still draw
a few quantitative comparisons.  For instance, Tyree~\cite{tyree-1983}
performed experiments on excised maple branches that absorbed water at a
maximum rate of 12\3cm${}^3$/h; for similar sized branches, our model
yields a comparable maximum absorption rate of roughly 13\3cm${}^3$/h as
well as qualitatively similar solution profiles. Another experiment by
Johnson \etal~\cite{johnson-tyree-dixon-1987} yielded total root uptake
of 2.0\3cm${}^3$ during freezing, followed by a much smaller uptake of
0.1\3cm${}^3$ during a subsequent freezing event.  We see similar
qualitative behaviour in our simulations, as well as measuring 2.2 and
0.2\3cm${}^3$ of water absorbed during the first and second freeze,
respectively.


\subsection{Two crucial mechanisms: Root water uptake and FPD} 
\begin{figure}
  \centering
  \begin{tabular}{lll}
    \figlabel{a} & & \myrevision{\figlabel{b}} \\
    \includegraphics[width=0.45\textwidth]{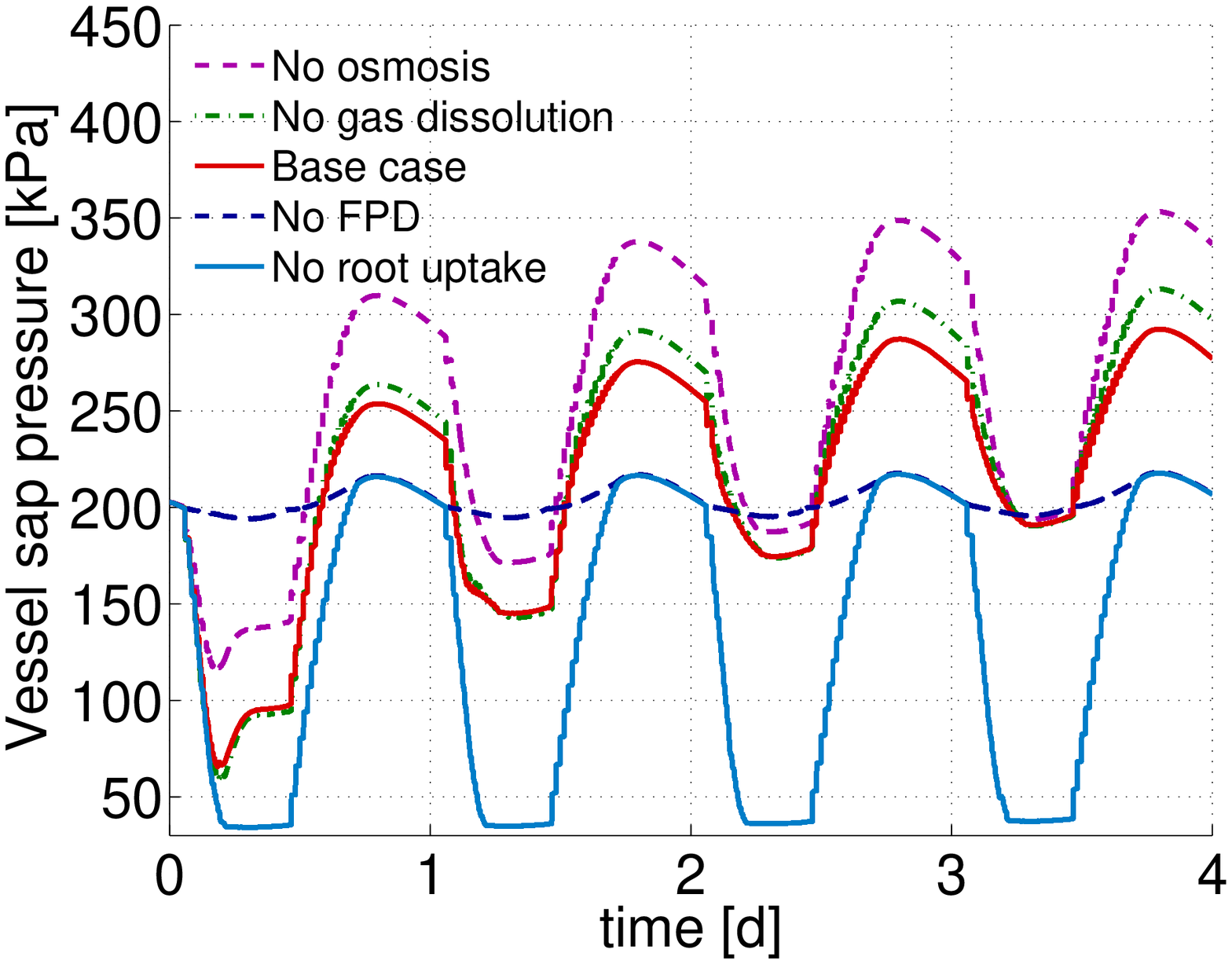}
    & & 
    \includegraphics[width=0.45\textwidth]{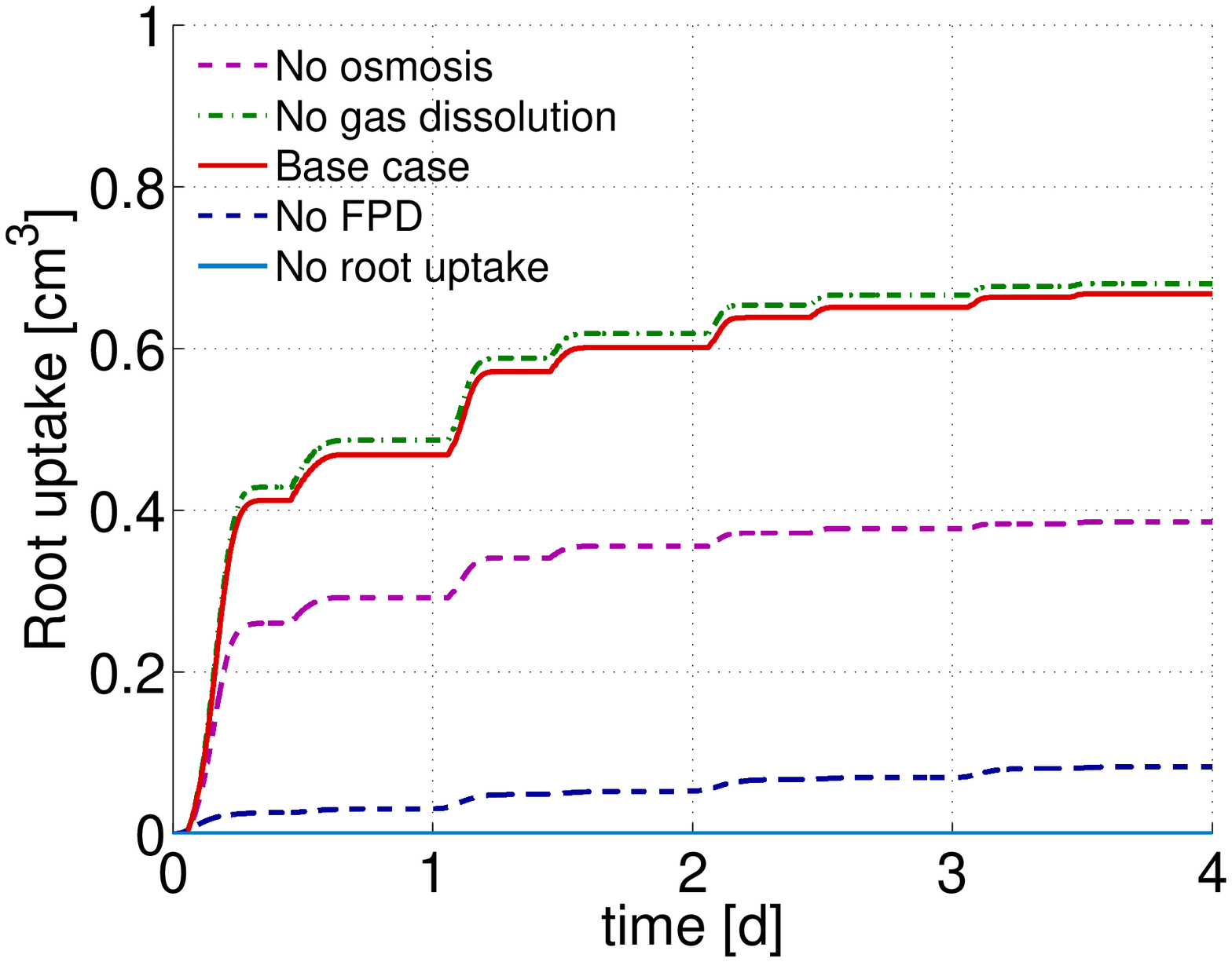}
  \end{tabular}
  \caption{{\bf Comparison of various physical mechanisms.}
    \figlabel{a} An investigation of the relative importance of various
    physical effects, depicting pressure with each of the following
    mechanisms turned off: FPD, root water uptake, osmosis, gas
    dissolution.  The `base case' and `no root uptake' curves are
    repeated from figure~\ref{fig:base}a for easy comparison.
    \myrevision{\figlabel{b} Corresponding curves for root uptake.}}
  \label{fig:mechanisms}
\end{figure}

\begin{figure}
  \centering
  \begin{tabular}{lll}
    \figlabel{a} & & 
    \figlabel{b} \\
    \includegraphics[width=0.45\textwidth]{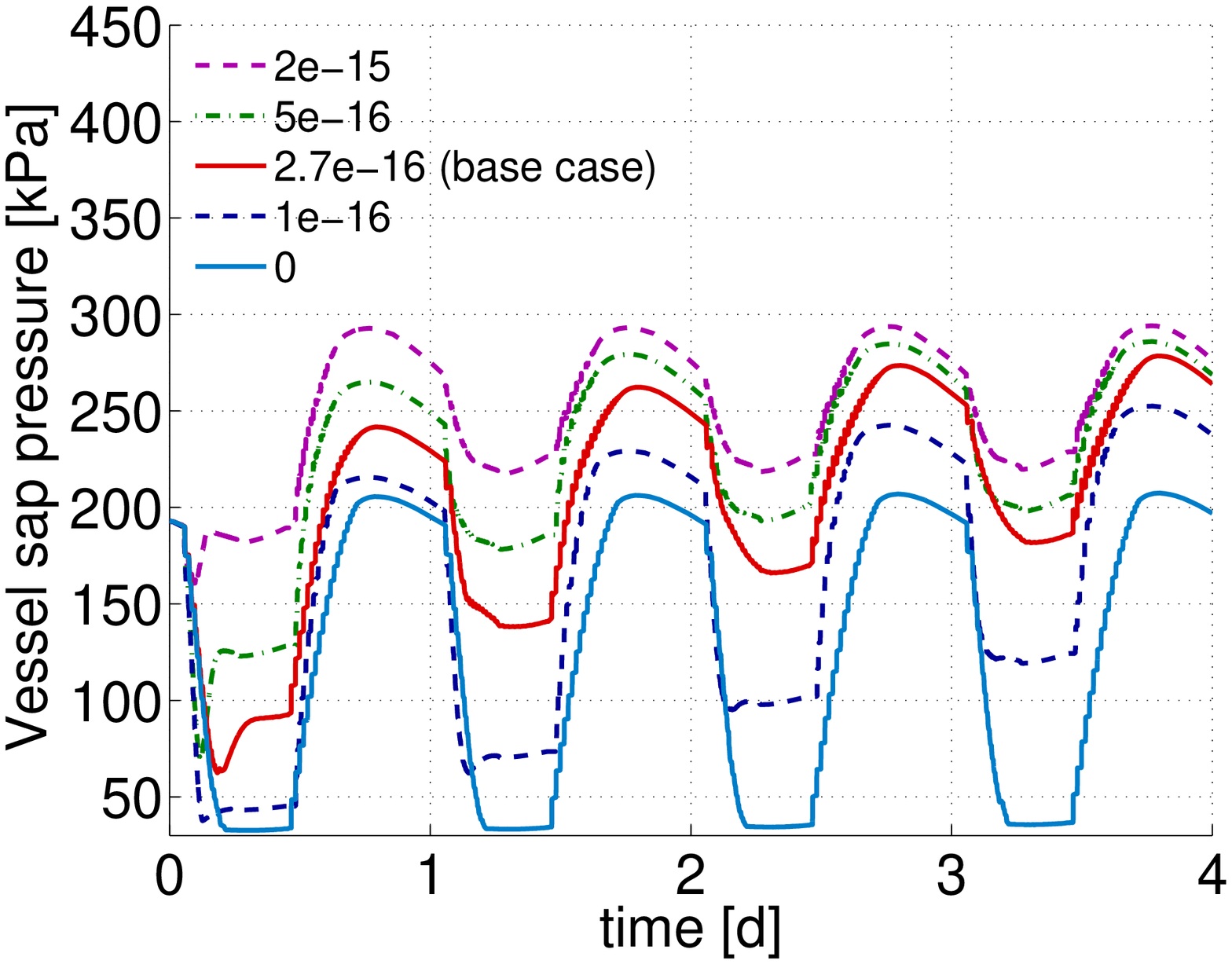}
    & & 
    \includegraphics[width=0.45\textwidth]{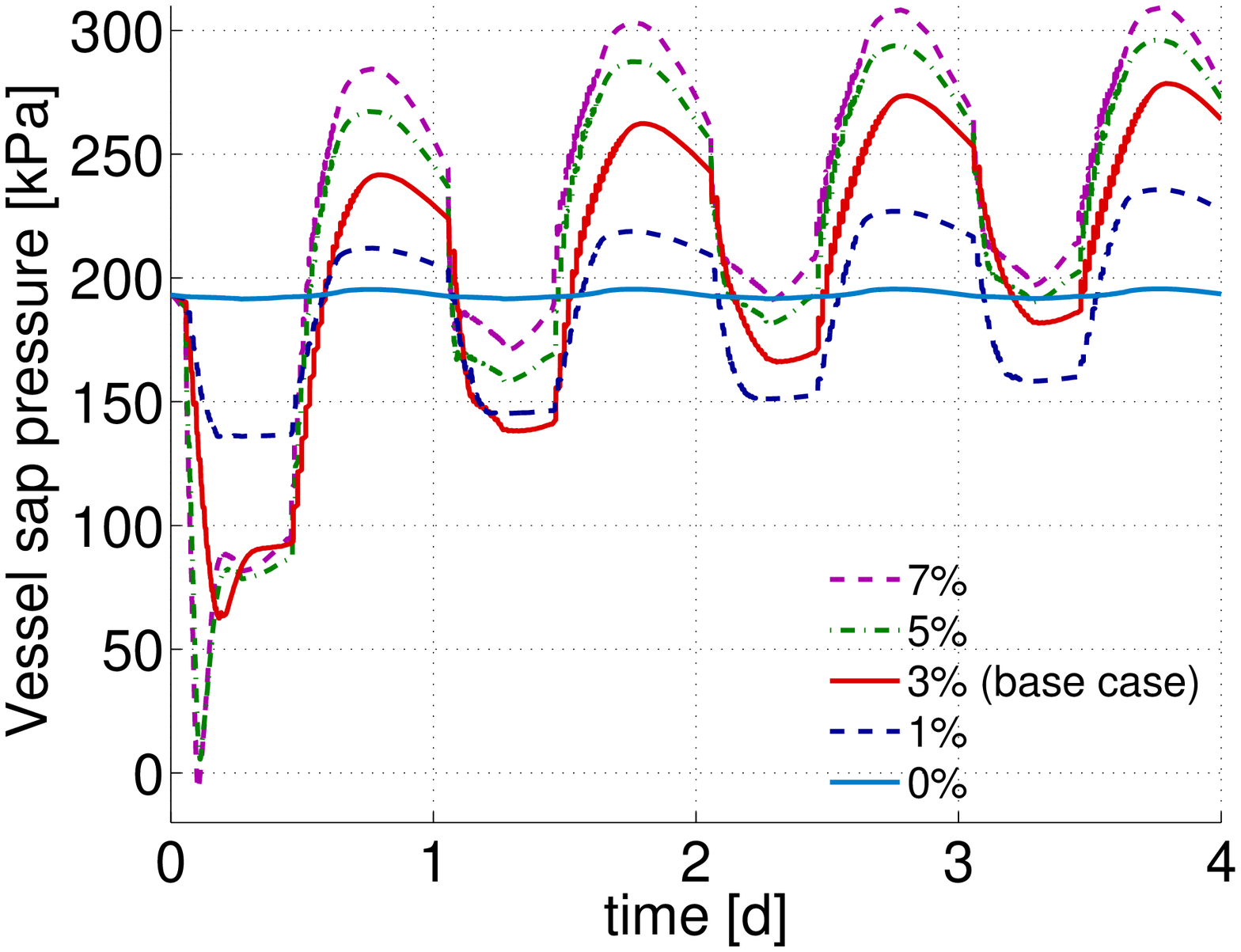}
  \end{tabular}
  \caption{{\bf Sensitivity of exudation pressure to parameters.} 
    \figlabel{a} Root hydraulic conductivity, $\hydcond_r$, in
    m\3s${}^{-1}$\3Pa${}^{-1}$. \figlabel{b}
    Sugar content in \%. 
  }
  \label{fig:vary-params}
\end{figure}

%
To evaluate the relative importance of the various physical mechanisms,
we present in figure~\ref{fig:mechanisms} the base case pressure
\myrevision{and root water uptake} alongside simulations with each of
the following mechanisms `turned off': FPD, root uptake (repeated from
figure~\ref{fig:base}a), osmosis and gas dissolution.  The first two
mechanisms clearly have the greatest impact on the build-up of exudation
pressure.  We already discussed the crucial role of root uptake in
facilitating pressure accumulation over multiple freeze--thaw cycles.
This effect is underscored by the plots in figure~\ref{fig:vary-params}a
depicting the pressure response when root conductivity $\hydcond_r$
varies between zero (no uptake) and nearly ten times the base value.

Without FPD, the vessel pressure remains nearly constant and there is
minimal root uptake, whereas without osmosis the vessel pressure
increases.  We therefore conclude that the predominant impact of sugar
on exudation is through FPD rather than osmosis, and even though $\Delta
T_{fpd}$ is small it nonetheless plays a critical role in facilitating
pressure transfer between fiber and vessel.  This dependence is
illustrated further by figure~\ref{fig:vary-params}b, where sugar
content is varied between 0 and 7\%\ \myrevision{and we observe that
  both net pressure build-up and oscillation amplitude increase with
  sugar content.}

\begin{figure}
  \centering
  \myrevision{
  \begin{tabular}{lll}
    \figlabel{a} & & \myrevision{\figlabel{b}} \\
    \includegraphics[width=0.45\textwidth]{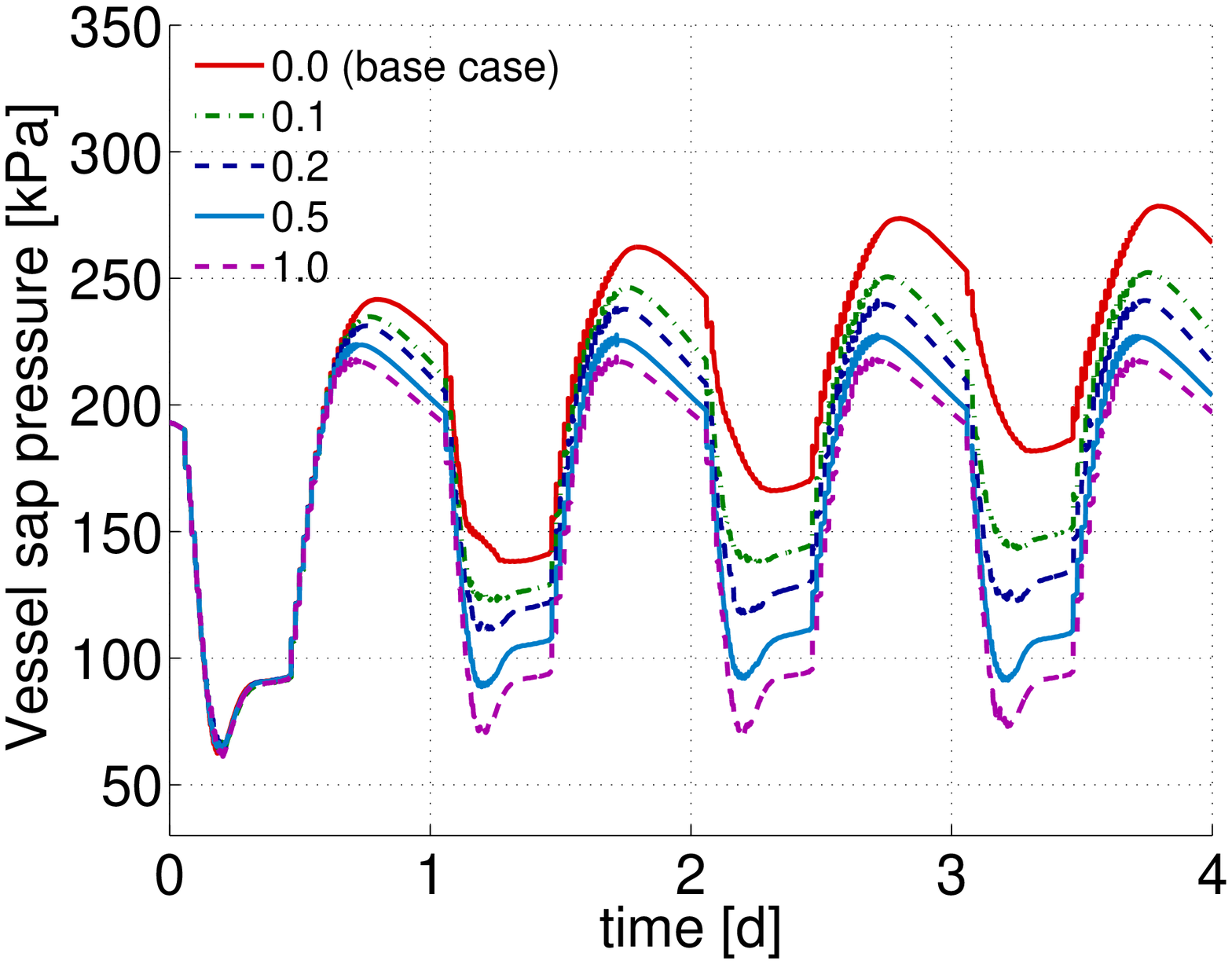}
    & & 
    \includegraphics[width=0.45\textwidth]{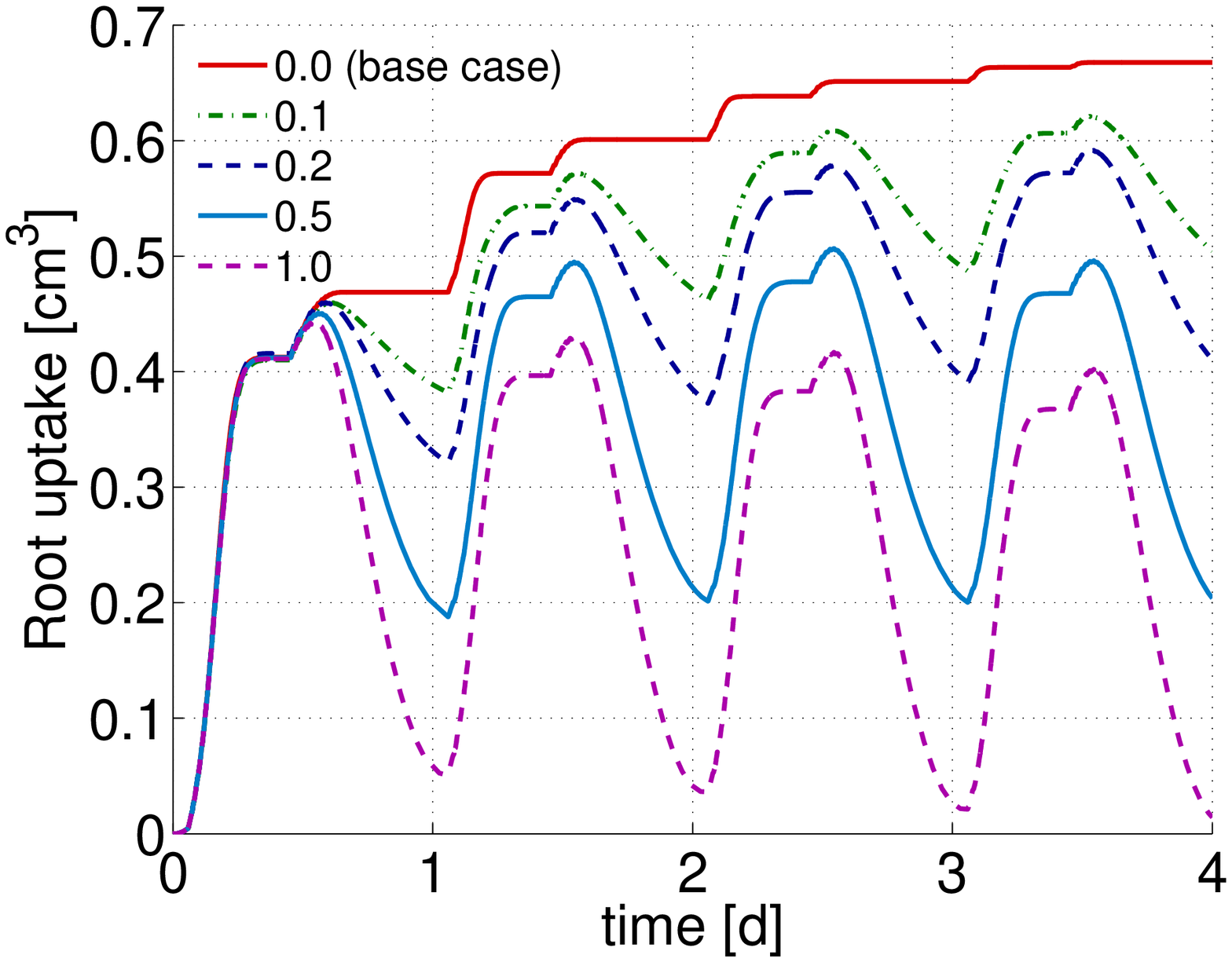}
  \end{tabular}
  }
  \caption{\myrevision{{\bf Sensitivity to root outflow.}  \figlabel{a}
      Pressure curves with non-zero conductivity to root outflow, where
      the ratio of outflow-to-inflow ratio varies between 0 and 1.
      \figlabel{b} Corresponding plots of root uptake.}}
  \label{fig:root-outflux}
\end{figure}

\myrevision{One assumption requiring further investigation is that of
  zero conductivity to root outflow in \eqref{eq:micro-Uroot}, which we
  motivated by citing experimental results that exhibit a small but
  still nonzero root outflow~\cite{henzler-etal-1999}.  To study this
  outflow effect, we take four different outflow conductivities equal to
  the inflow value $\hydcond_r$ scaled by a factor between 0 and 1
  (where 0 corresponds to the base case).  The results in
  figure~\ref{fig:root-outflux} clearly show that allowing even a small
  outflow has a major influence on the root water uptake by preventing
  accumulation of water over multiple freeze--thaw cycles and thereby
  reducing build-up of exudation pressure.  Because of the obvious
  sensitivity of these results to root outflow, a more extensive
  experimental study of root conductivity in maple is warranted.}

\myrevision{We end this section by addressing the seemingly
  counter-intuitive result in figure~\ref{fig:mechanisms}a that
  introducing osmosis decreases vessel sap pressure.  This result can be
  most easily explained by considering the water flux equation
  \eqref{eq:micro-U} over a long enough time that the fiber and vessel
  have reached a quasi-steady state and $\partial_t U\approx 0$.  Then
  \eqref{eq:micro-U} reduces to the simple pressure balance
  \begin{gather*}
    \underbrace{\left( p_g^v - \frac{2\sigma_{gw}}{r} \right)}_{p_w^v} - 
    \underbrace{\left( p_g^f - \frac{2\sigma_{gw}}{s_g} \right)}_{p_w^f} 
    - \, p_{osm} + p_i^f \approx 0.
  \end{gather*}
  The ice--water capillary pressure $p_i^f$ is a constant, and our
  simulations show that osmosis has relatively small impact on fiber
  bubble size and pressure (the latter effect was discussed in
  \cite{ceseri-stockie-2013}).  Therefore, the primary influence of
  osmosis is within the vessel: osmotically-driven flow from fiber to
  vessel compresses the vessel bubble which not only increases the
  vessel gas pressure $p^v_g$, but also increases the capillary pressure
  term (via a reduction in bubble radius $r$).  The contribution from
  surface tension dominates and so the net effect is actually a
  \emph{decrease} in vessel sap pressure $p^v_w$, which is consistent
  with figure~\ref{fig:mechanisms}a and the results reported
  in~\cite{ceseri-stockie-2013}.}

\subsection{Phase change dynamics on the microscale} 
%
When a completely frozen tree warms above 0\3\degC\ during the day, a
thawing front develops near the bark (wherein water and sap are frozen
ahead of the front and thawed behind) and advances into the stem; an
analogous scenario occurs upon freezing.  Clearly, the `interesting'
solution dynamics will occur in the vicinity of this front, and hence
knowledge of phase change on the microscale is desirable for
understanding solution behaviour.  Over a century ago,
Wiegand~\cite{wiegand-1906} recognized the existence of freezing and
thawing fronts that `penetrate the wood in a wave-like manner' and in
which `but few cells would actually take part in the production of
pressure at any one time'; however, there has so far been no attempt to
develop a mathematical model for this phenomenon.  In particular, the
role of FPD in governing the progress of these phase transitions
throughout the sapwood has not been investigated before.


\begin{figure}
  \centering
  \begin{tabular}{lll}
    \figlabel{a} & & \figlabel{b} \\[-0.1cm]
    \includegraphics[width=0.45\textwidth]{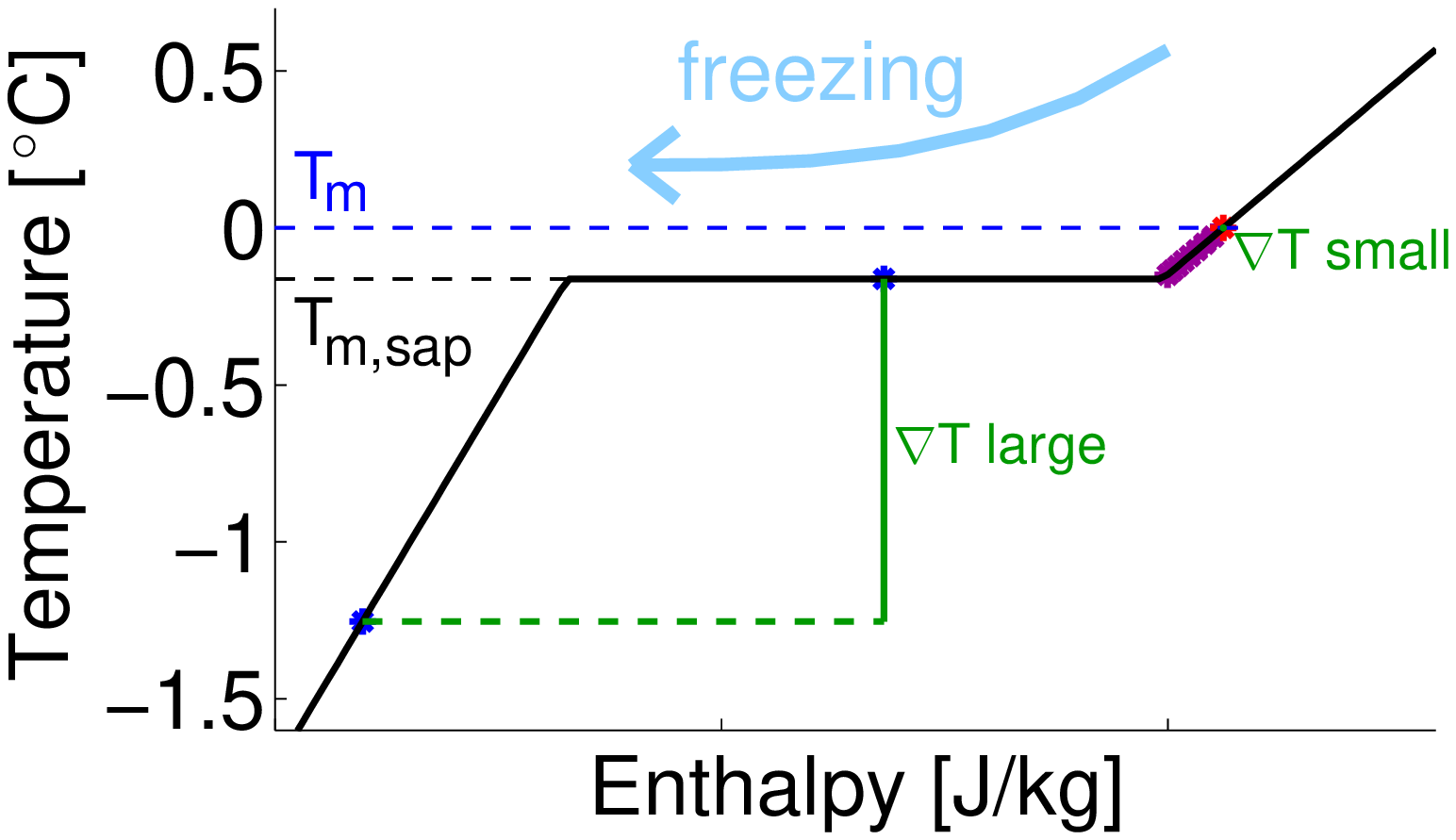}
    & &
    \includegraphics[width=0.45\textwidth]{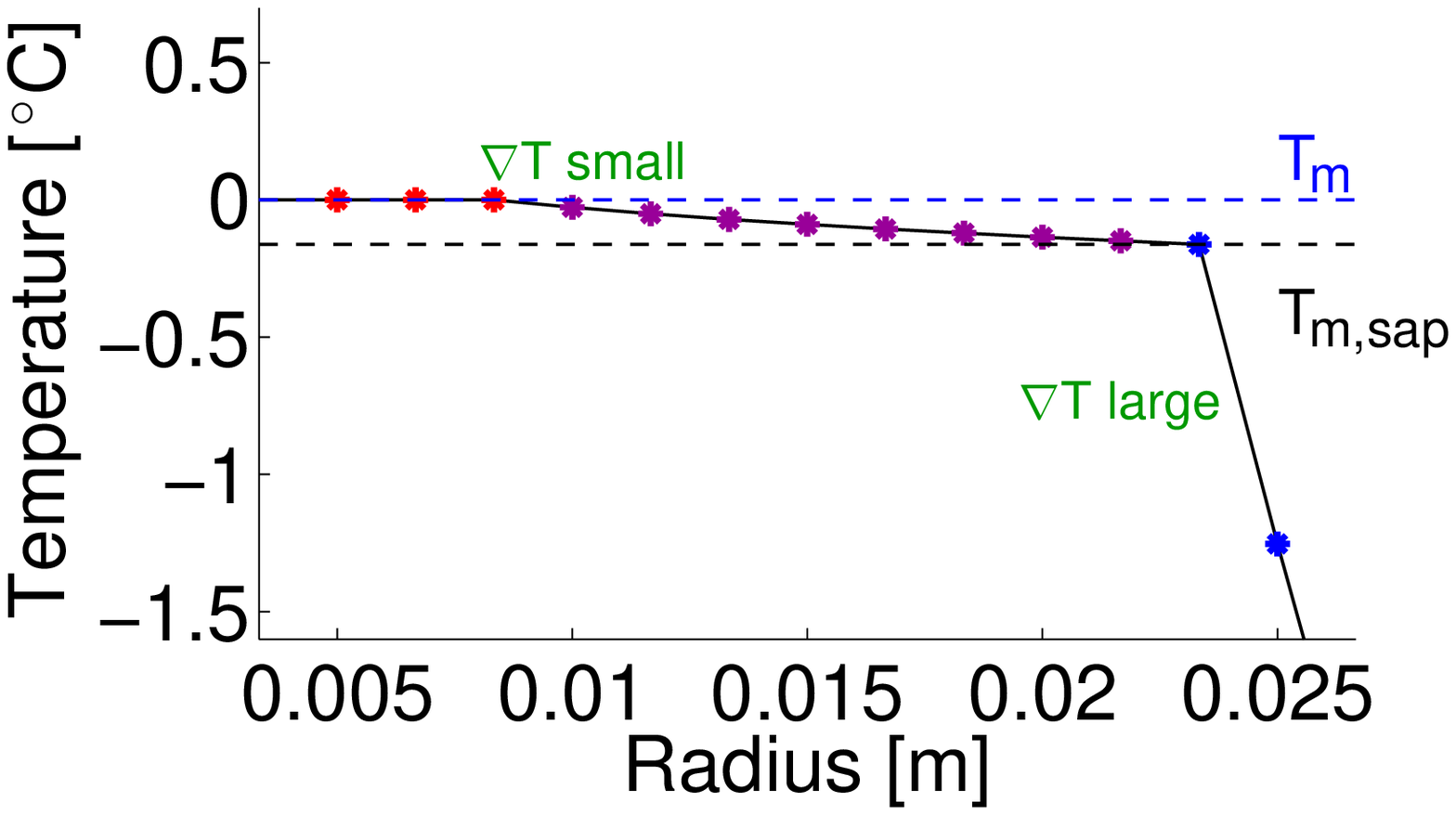}
    \\
    \figlabel{c} & & \figlabel{d} \\[-0.1cm]
    \includegraphics[width=0.45\textwidth]{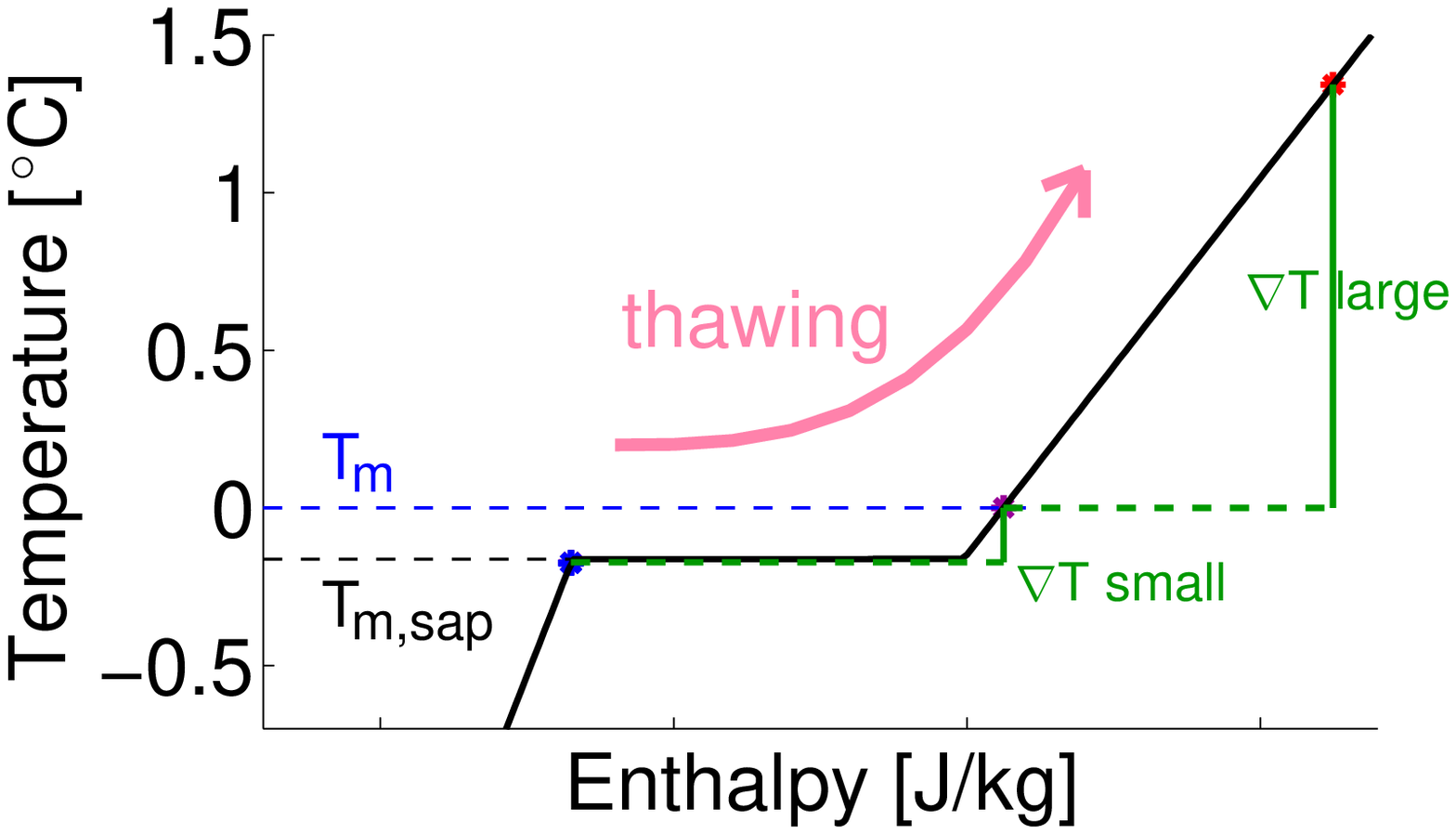}
    & &
    \includegraphics[width=0.45\textwidth]{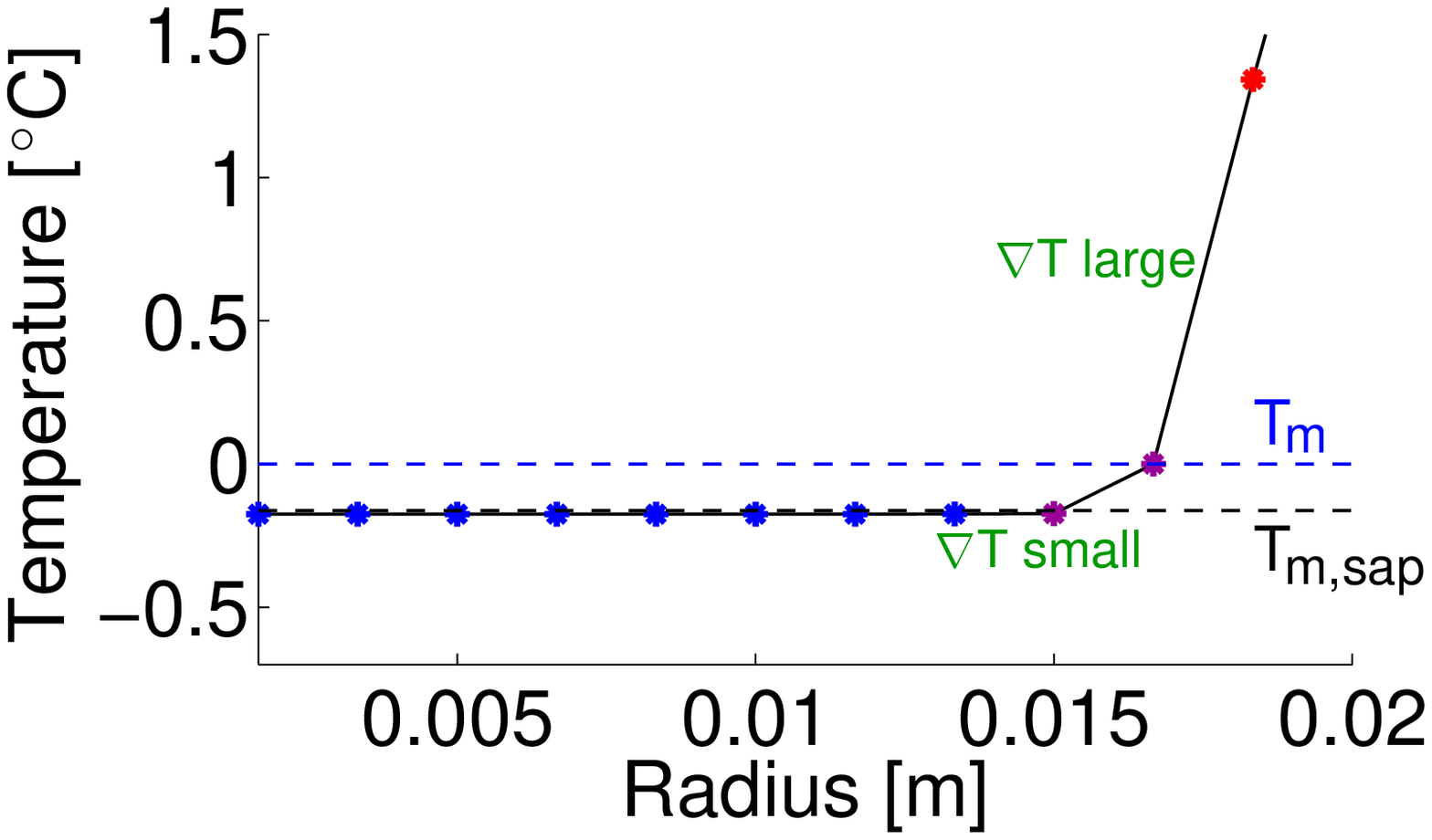}
  \end{tabular}
  \caption{{\bf Local phase change dynamics.}  Plots of temperature
    versus radius and enthalpy for a fixed time in the middle of a
    freezing event (a,b, top) and a thawing event (c,d, bottom).  Points
    correspond to discrete solution values on an equally-spaced radial
    grid and are coloured according to the current state of fiber and
    vessel: blue if both are frozen; red if both are thawed; purple if
    fiber is frozen and vessel is thawed.}
  \label{fig:enthalpy}
\end{figure}

Phase change dynamics are most easily studied by means of a
temperature--enthalpy diagram as depicted in
figures~\ref{fig:enthalpy}a,c, which are each taken at a fixed time
during a freeze or thaw event.  Both plots feature a plateau region at
the \myrevision{melting} temperature, which has a horizontal extent
equal to the \emph{enthalpy of fusion}.  Note that there are two
distinct \myrevision{melting} temperatures in fiber and vessel equal to
$T_m$ and $T_{m,sap}=T_m-\Delta T_{fpd}$ respectively.  The
corresponding plots of temperature versus radius are shown in
figures~\ref{fig:enthalpy}b,d which depict the local state of each point
within the tree stem.  For example, in the freezing case (top) the three
grid points closest to the stem centre are completely thawed (red), the
outermost point is frozen (blue), and the intervening points are
undergoing freezing (purple).  Owing to FPD, water in the fiber freezes
before the vessel sap, thereby introducing a time delay in formation of
ice between the fiber and vessel.

For both freezing and thawing, the bulk of the stem is in a state
located at the leading edge of the enthalpy plateau (right edge for
freezing, left edge for thawing).  This behaviour can be explained by
considering the local rate of phase change: conservation of energy at a
phase interface is expressed mathematically using the well-known
{Stefan condition}, which states that the rate of freezing (or
thawing) is proportional to the temperature gradient.  Referring to
figures~\ref{fig:enthalpy}b,d, the temperature difference between
adjacent points is smaller near the tree centre and larger near
the bark.
Consequently, at any location in the tree a freezing event begins
within the fiber as a slow process, followed at a later time in the
vessel which freezes relatively quickly.  In contrast, a thawing event
begins with a slow thawing of the vessel sap, followed by rapid thawing
in the fiber.

\section{Concluding remarks}
\label{sec:conclude}

We have developed the first complete mathematical model for the tree sap
exudation process based on a prevailing freeze--thaw hypothesis.  We
introduced a number of \myrevision{additions} to this hypothesis, and
identified root water uptake and freezing point depression (FPD) as the
two main driving mechanisms for sap exudation.  \myrevision{In
  particular, we showed that the primary mechanism whereby sugar induces
  exudation pressure is via the FPD and not osmosis as was previously
  believed.}  Numerical simulations of the governing equations
demonstrate qualitative and quantitative agreement with experimental
data on sugar maple and the related species black walnut.  The quality
of agreement is striking considering that the model parameters were
determined using a minimum of parameter fitting.  Our work clearly
demonstrates the need for further experiments on sugar maple that
parallel the work of Am\'eglio \etal\ on
walnut~\cite{ameglio-etal-2001}.  \myrevision{Our model results lead to
  the important conclusion that FPD is a primary driver of sap
  exudation, which also requires experimental validation.}  Furthermore,
because we have only rough estimates at present for two of the model
inputs -- capillary pore size $r_{cap}$ and root conductivity
$\hydcond_r$ \myrevision{(especially differences between conductivity to
  inflow and outflow)} -- more accurate measurements of these parameters
are also required.

Our model provides an ideal platform from which to investigate 
other problems related to sap flow in maple and related species.
First of all, we aim to extend our current model of a 2D stem
cross-section to three dimensions.  This will permit us to incorporate
variations in gravitational pressure head and sugar concentration with
height~\cite{wiegand-1906} and to study problems of practical importance
to the maple syrup industry such as optimizing tap-hole placement or
determining sensitivity to changes in soil or climatic conditions.
Finally, there are a number of intriguing parallels between exudation
and the phenomenon of freeze-induced winter
embolism~\cite{sperry-etal-1988, yang-tyree-1992}
that are also worthy of future investigation.

\begin{competingint}
  We have no competing interests.
\end{competingint}

\begin{authorcontrib}
  JMS designed the study.  IG and MC designed the numerical
  algorithm. IG carried out the computational studies.  All authors
  derived the mathematical model, analyzed the results, and wrote the
  manuscript.  All authors gave final approval for publication.
\end{authorcontrib}

\myrevision{
\begin{acknowledgements}
  We are indebted to Chris Budd (University of Bath) for his
  helpful comments on an earlier version of this manuscript.  
\end{acknowledgements}
}

\begin{fundingstmt}
  This work was supported by research grants from the Natural Sciences
  and Engineering Research Council of Canada and the North American
  Maple Syrup Council (to JMS), a Postdoctoral Fellowship from Mitacs
  (to MC) and a Feodor Lynen Fellowship from the Alexander von
  Humboldt Stiftung (to IG).
\end{fundingstmt}


\providecommand{\noopsort}[1]{}

\label{lastpage}
\end{document}